\documentclass[12pt,preprint]{aastex}
\usepackage{amsmath,amssymb}
\usepackage{graphicx}
\usepackage{float}
\usepackage{multirow}

\slugcomment{KSUPT-12/5 \hspace{0.5truecm} November 2012}

\begin{document}
\title{Hubble parameter measurement constraints on dark energy}
\author{Omer Farooq\altaffilmark{1}, Data Mania\altaffilmark{1,2}, and Bharat Ratra\altaffilmark{1}}

\altaffiltext{1}{Department of Physics, Kansas State University, 
                 116 Cardwell Hall, Manhattan, KS 66506, USA \ 
                 omer@phys.ksu.edu, mania@phys.ksu.edu, and ratra@phys.ksu.edu}
\altaffiltext{2}{Center for Elementary Particle Physics, 
                 Ilia State University, 3-5 Cholokashvili Ave., 
                 Tbilisi 0179, Georgia}

\begin{abstract}

 We use 21 Hubble parameter versus redshift data points, from \cite{simon05},
\cite{gaztanaga09}, \cite{Stern2010}, and \cite{moresco12}, to
place constraints on model parameters of constant and time-evolving
dark energy cosmologies. The inclusion of the 8 new \cite{moresco12}
measurements results in $H(z)$ constraints more restrictive than 
those derived by \cite{Chen2011b}. These constraints are now almost 
as restrictive as those that follow from current Type Ia supernova 
(SNIa) apparent magnitude versus redshift data \citep{suzuki12},
which now more carefully account for systematic uncertainties. 
This is a remarkable result. We emphasize however that SNIa data have 
been studied for a longer time than the $H(z)$ data, possibly 
resulting in a better estimate of potential systematic errors in the 
SNIa case. A joint analysis of the $H(z)$, baryon acoustic 
oscillation peak length scale, and SNIa data favors a spatially-flat 
cosmological model currently dominated by a time-independent 
cosmological constant but does not exclude slowly-evolving dark energy.

\end{abstract}

\maketitle
\section{Introduction}

The expansion rate of the Universe changes with time, initially slowing
when matter dominated, because of the mutual gravitational attraction of 
all the matter in it, and more recently accelerating. A number of 
cosmological observations now strongly support the idea that the
Universe is spatially flat (provided the dark energy density is close to
or time independent) and is currently undergoing accelerated cosmological 
expansion. A majority of cosmologists consider dark energy to be
the cause of this observed accelerated cosmological 
expansion.\footnote{
Some cosmologists instead view these observations as an indication 
that general relativity needs to be modified on these large 
length scales. For recent reviews of modified gravity see \cite{Tsujikawa2010},
\cite{Bolotin2011}, \cite{Capozziello2011}, \cite{Starkman2011}, and
references therein. In this paper we assume that general relativity 
provides an adequate description of gravitation on cosmological scales.}
This dark energy, most simply thought of as a negative pressure substance,  
dominates the current cosmological energy budget. For reviews of dark energy
see \cite{Bass2011}, \cite{Jimenez2011}, \cite{Li2011a}, \cite{Bolotin2011}, 
and references therein. 

Three observational techniques provide the strongest evidence
for dark energy: SNIa apparent magnitude measurements as a function of 
redshift \citep[e.g.,][]{Sullivan2011, suzuki12, Li2011b, Barreira2011};
cosmic microwave background (CMB) anisotropy data 
\citep[][and references therein]{Podariu2001b, Komatsu2011} combined with low 
estimates of the cosmological mass density \citep[e.g.,][]{chen03b}, provided 
the dark energy density is close to or time independent; and baryon acoustic 
oscillation (BAO) peak length scale measurements 
\citep[e.g.,][]{beutler2011, blake11, Mehta2012}.

The ``standard" model of cosmology is the spatially-flat $\Lambda$CDM model
\citep{peebles84}. In this model about $73\%$ of the current 
energy budget is dark energy, Einstein's cosmological constant $\Lambda$. 
Non-relativistic cold dark matter (CDM) is the next largest contributer to 
the energy budget (around $23\%$), followed by non-relativistic 
baryonic matter (about $5\%$). For reviews of the standard model of cosmology
see \cite{Ratra08} and references therein. It has been known for a while 
that the $\Lambda$CDM model is reasonably consistent with most observations
\citep[see, e.g.,][for early indications]{jassal10, wilson06, Davis2007,
allen08}.\footnote{
Note, however, that the ``standard'' CDM structure formation model, which
is assumed in the $\Lambda$CDM model, might need modification 
\citep[see][and references therein]{Peebles&Ratra2003, 
Perivolaropoulos2010}.} 
In the $\Lambda$CDM model the dark energy density is constant in time and does 
not vary in space.

Although most predictions of the $\Lambda$CDM model are reasonably 
consistent with the measurements, the $\Lambda$CDM model has some curious
features. For instance, the measured cosmological constant energy 
density is 120 orders of magnitude smaller than the energy density 
naively expected from quantum field theory (this is known as the fine-tuning 
puzzle). A second curiosity is what is known as the coincidence 
puzzle: the energy density of a cosmological constant, $\rho_\Lambda$, is 
independent of time, but that of matter, $\rho_{\rm m}$, decreases with 
time during the cosmological expansion, so it is curious why we (observers) 
happen to live at this (apparently) special time, when the dark energy and 
the non-relativistic matter energy densities are of comparable 
magnitude.

These puzzles could be partially resolved if the dark energy 
density is a slowly decreasing function of time \citep[][]{Peebles&Ratra1988,
Ratra&Peebles1988}. In this case the dark energy density will remain 
comparable to the non-relativistic matter density for a longer time. 
For recent discussions of time-varying dark energy models, see
\cite{Bauer2011}, \cite{Chimento2011}, \cite{Granda2011}, 
\cite{GarciaBellido2011}, \cite{Basilakos2012}, \cite{Sheykhi2012}, 
\cite{Brax2012}, \cite{Hollenstein2012}, \cite{Cai2012}, and references 
therein. In this paper we will consider two dark energy models (with
dark energy being either a cosmological constant or a slowly-evolving 
scalar field $\phi$) as well as a dark energy parameterization.

In the $\Lambda$CDM model, time-independent dark energy density (the 
cosmological constant $\Lambda$) is modeled as a spatially homogeneous 
fluid with equation of state $p_{\rm \Lambda} = -\rho_{\rm \Lambda}$. 
Here $p_{\rm \Lambda}$ and $\rho_{\rm \Lambda}$ are the fluid pressure 
and energy density. In describing slowly-decreasing dark 
energy density much use has been made of a parameterization known as XCDM.
Here dark energy is modeled as a spatially homogeneous $X$-fluid with 
equation of state $p_{\rm X}=w_{\rm X}\rho_{\rm X}$. The equation of 
state parameter $w_{\rm X}<-1/3$ is independent of time and $p_{\rm X}$ 
and $\rho_{\rm X}$ are the  pressure and energy density of the $X$-fluid. 
When $w_{\rm X}=-1$ the XCDM parameterization reduces 
to the complete and consistent $\Lambda$CDM model. For any other 
value of $w_{\rm X}<-1/3$ the XCDM parameterization is incomplete as it 
cannot describe spatial inhomogeneities \citep[see, e.g.][]{ratra91, 
podariu2000}. For computational simplicity, in the 
XCDM case we assume a spatially-flat cosmological model.  

The $\phi$CDM model is the simplest, consistent and complete model of
slowly-decreasing dark energy density \citep{Peebles&Ratra1988, 
Ratra&Peebles1988}. In this model dark energy is modeled as a scalar 
field, $\phi$, with a gradually decreasing (in $\phi$) potential energy 
density $V(\phi)$. Here we assume an inverse power-law potential energy 
density $V(\phi) \propto \phi^{-\alpha}$, where $\alpha$ is a nonnegative 
constant \citep{Peebles&Ratra1988}. When $\alpha = 0$ the
$\phi$CDM model reduces to the corresponding $\Lambda$CDM case. 
For computational simplicity, we again only consider the spatially-flat
cosmological case for $\phi$CDM. 

As mentioned above, for some time now, most observational constraints have 
been reasonably consistent with the predictions of the ``standard'' 
spatially-flat 
$\Lambda$CDM model. CMB anisotropy, SNIa, and BAO measurements
provide the strongest support for this conclusion. However, the
error bars associated with these three types of data are still too 
large to allow for a significant observational discrimination between 
the $\Lambda$CDM model and the two simple time-varying dark energy models 
discussed above. This is one motivation for considering additional kinds 
of data.

If the constraints from the new data differ considerably from the old 
ones, this could mean that at least one of the data sets had an 
undetected systematic error, or it could mean that the model being
tested is observationally inconsistent. Either of these is an important 
result. On the other hand, if the constraints from the new and the old 
data are consistent, then a joint analysis of all the data could result in 
tighter constraints, and so might result in significantly discriminating 
between constant and time-varying dark energy models.

Other measurements that have been used to constrain cosmological 
parameters\footnote{
For reviews see \cite{Albrecht2006}, \cite{Weinberg2012}, and references 
therein.}
include galaxy cluster gas mass fraction as a function of redshift 
\citep[e.g.,][]{allen08, Samushia&Ratra2008, ettori09, tong11, Lu2011}, 
galaxy cluster and other large-scale structure properties 
\citep[][and references therein]{campanelli11, deboni10, mortonson2011, 
Devi2011, Wang2012}, gamma-ray burst luminosity distance as a function of 
redshift \citep[e.g.,][]{Samushia&Ratra2010, Wang2011, Busti2012}, 
lookback time as a function of redshift \citep[][and references 
therein]{Samushiaetal2010, dantas11, Tonoiu2011}, HII starburst galaxy 
apparent magnitude as a function of redshift \citep[e.g.,][]{plionisetal10, 
plionisetal11, Mania2012}, angular size as a function of redshift 
\citep[e.g.,][]{guerra00, Bonamente2006, Chen2012}, and strong 
gravitational lensing \citep[][and references therein]{chae04, lee07, 
biesiada10, Zhang2010}.\footnote{
Future space-based SNIa and BAO-like meassurements 
\citep[e.g.,][]{podariu01a, Samushia2011, Sartoris2012, Basse2012, Pavlov2012},
as well as measurements based on new techniques 
\citep[][and references therein]{Jennings2011, vandeWeygaert2011, 
Ziaeepour2012} should soon provide interesting constraints on cosmological
parameters.}
Of particular interest to us here are measurements of the Hubble parameter
as a function of redshift \citep[e.g.,][]{Jimenezetal2003, Samushia&Ratra2006,
samushia07, Sen&Scherrer2008, Panetal2010, Chen2011b, Kumar2012, 
WangZhang2012, Duan2012, bilicki12, seikel12}. While the constraints from 
these data are typically less restrictive than
those derived from the SNIa, CMB anisotropy, and BAO data, both types of
measurements result in largely compatible constraints that generally 
support a currently accelerating cosmological expansion. This provides
confidence that the broad outlines of a ``standard'' cosmological
model are now in place. 

\begin{table}
\begin{center}
\begin{tabular}{cccc}
\hline\hline
~~$z$ & ~~$H(z)$ &~~~~~~~ $\sigma_{H}$ &~~ Reference\\
~~~~~    & (km s$^{-1}$ Mpc $^{-1}$) &~~~~~~~ (km s$^{-1}$ Mpc $^{-1}$)& \\
\tableline

0.090&~~	69&~~~~~~~	12&~~	1\\
0.170&~~	83&~~~~~~~	8&~~	1\\
0.179&~~	75&~~~~~~~	4&~~	4\\
0.199&~~	75&~~~~~~~	5&~~	4\\
0.240&~~	79.69&~~~~~~~	2.65&~~	2\\
0.270&~~	77&~~~~~~~	14&~~	1\\
0.352&~~	83&~~~~~~~	14&~~	4\\
0.400&~~	95&~~~~~~~	17&~~	1\\
0.430&~~	86.45&~~~~~~~	3.68&~~	2\\
0.480&~~	97&~~~~~~~	62&~~	3\\
0.593&~~	104&~~~~~~~	13&~~	4\\
0.680&~~	92&~~~~~~~	8&~~	4\\
0.781&~~	105&~~~~~~~	12&~~	4\\
0.875&~~	125&~~~~~~~	17&~~	4\\
0.880&~~	90&~~~~~~~	40&~~	3\\
0.900&~~	117&~~~~~~~	23&~~	1\\
1.037&~~	154&~~~~~~~	20&~~	4\\
1.300&~~	168&~~~~~~~	17&~~	1\\
1.430&~~	177&~~~~~~~	18&~~	1\\
1.530&~~	140&~~~~~~~	14&~~	1\\
1.750&~~	202&~~~~~~~	40&~~	1\\
\hline\hline
\end{tabular}
\end{center}
\caption{Hubble parameter versus redshift data. Last column reference numbers:
1. \cite{simon05}, 2. \cite{gaztanaga09}, 3. \cite{Stern2010}, 
4. \cite{moresco12}.
}\label{tab:Hz}
\end{table}

In this paper we use the 21 $H(z)$ measurements of \cite{simon05}, 
\cite{gaztanaga09}, 
\cite{Stern2010}, and  \cite{moresco12} (listed in Table 1)\footnote{ 
We do not include the 4 recent \cite{Zhang2012} $H(z)$ measurements as they 
have somewhat larger error bars and do not affect our results.}  
to constrain the $\Lambda$CDM and  $\phi$CDM models and the XCDM 
parametrization. The inclusion of the 8 new \cite{moresco12} measurements 
(with smaller error bars compared to the earlier data) in the analysis 
results in tighter constraints than those recently derived by 
\cite{Chen2011b} from the previous largest set of $H(z)$ measurements
considered. The new $H(z)$ data constraints derived here are compatible 
with cosmological parameter constraints determined by other techniques. 
For the first time, these
$H(z)$ limits are almost as constraining as those derived from the most 
recent SNIa data compilation of \cite{suzuki12}. In addition to the 
tighter $H(z)$ limits resulting from the new data, this is partially 
also a consequence of the fact that a more careful analysis of the 
SNIa measurements \citep{suzuki12} has 
resulted in a larger systematic error estimate and thus weaker SNIa
constraints. We emphasize that the study of $H(z)$ data is much less 
mature than that of SNIa apparent magnitude data, so there is the
possibility that future $H(z)$ error bars might be larger than what 
we have used in our analysis here. In addition to deriving 
$H(z)$-data only constraints, we also use these $H(z)$ data in 
combination with recent BAO and SNIa measurements to jointly constrain 
cosmological parameters in these models.\footnote{
See \cite{Moresco2012b} and \cite{WangX2012} for analyses that use 
most of these $H(z)$ data in conjunction with CMB anisotropy and
other data to constrain cosmological parameters.}
Adding the $H(z)$ data tightens the constraints, somewhat significantly 
in some parts of parameter space for some of the models we study.

Our paper is organized as follows. In Sec.\ {\ref{equations}} we
present the basic equations of the three dark energy models we
consider. Constraints from the $H(z)$ data are derived in Sec.\
{\ref{HzData}}. In Sec.\ {\ref{SNeIa}} we determine constraints 
from recent SNIa apparent magnitude data. In Sec.\ {\ref{BAO}} we 
derive constraints from recent BAO data. Joint constraints on 
cosmological parameters, from a combined analysis of the three 
data sets, for the three models we consider, are presented in  
Sec.\ {\ref{Joint}}. We conclude in 
Sec.\ {\ref{summary}}.

\label{intro}

\section{Dark energy models}
\label{equations}

In this section we summarize properties of the two models 
($\Lambda$CDM and $\phi$CDM) and the one parametrization (XCDM)
we use in our analyses of the data.

To determine how the Hubble parameter $H(z)$ evolves in these 
models, we start from the Einstein equation of general relativity
\begin{equation}
R_{\mu\nu} - \frac{1}{2}g_{\mu\nu}R = 8 \pi G T_{\mu\nu} - \Lambda g_{\mu\nu}.
\end{equation} 
Here $g_{\mu\nu}$ is the metric tensor, $R_{\mu\nu}$ and $R$ are the Ricci 
tensor and scalar, $T_{\mu\nu}$ is the energy-momentum tensor 
of any matter present, $\Lambda$ is the cosmological constant, and 
$G$ is the Newtonian gravitational constant.

The energy-momentum tensor for an ideal fluid is $ T_{\mu\nu} = 
{\rm diag} (\rho, p, p, p)$, where $\rho$ is the energy density 
and $p$ the pressure. Assuming spatial homogeneity, the Einstein
equation reduces to the two independent Friedmann equations
\begin{equation}
\label{eq:F1}
\left(\frac{\dot{a}}{a}\right)^2 = \frac{8\pi G}{3} \rho +\frac{\Lambda}{3} - 
\frac{K^2}{a^2},
\end{equation}
\begin{equation}
\label{eq:F2}
\frac{\ddot{a}}{a} =-\frac{4\pi G}{3} (\rho +3p)+\frac{\Lambda}{3}.
\end{equation}
Here $a(t)$ is the cosmological scale factor, an overdot denotes a 
derivative with respect to time, and $K^2$ represents the curvature of 
the spatial hypersurfaces. 
These equations, in conjunction with the equation of state, 
\begin{equation}
\label{eq:EoS}
   p= p(\rho) = \omega \rho, 
\end{equation}
where $\omega$ is the dimensionless equation-of-state parameter (with
$\omega=-1$ corresponding to a cosmological constant and $\omega 
< -1/3$ corresponding to the XCDM parametrization), govern
the evolution of the scale factor and matter densities.

Taking the time derivative of Eq.\ (\ref{eq:F1}) and putting 
it in Eq.\ (\ref{eq:F2}) and then using Eq.\ (\ref{eq:EoS})
yields the energy conservation equation
\begin{equation}
\label{eq:EC}
   \dot{\rho} = -3 \frac{\dot{a}}{a} (\rho +p) 
   = -3 \rho \frac{\dot{a}}{a} (1+\omega)
\end{equation} 
For a non-relativistic gas (matter) $\omega=\omega_m =0$ and 
$\rho_m \propto a^{-3}$, 
and for a cosmological constant $\omega=\omega_\Lambda =-1$ and 
$\rho_\Lambda = \Lambda/(8\pi G)$= constant
($\dot{\rho_{\rm \Lambda}}$=0).
Solving Eq.\ (\ref{eq:EC}), the time-dependent energy density is
\begin{equation}
\label{eq:EC1}
   \rho(t)=\rho_0 \left(\frac{a_0}{a}\right)^{3(1+\omega)}
\end{equation}
where $\rho_0$ and $a_0$ are the current values of the fluid energy density 
and the scale factor. If there are a number of different species 
of non-interacting particles, then Eq.\ (\ref{eq:EC1}) holds separately
for each of them.

The ratio $\dot{a}(t)/a(t) = H(t)$  is called the Hubble parameter. 
The present value of the Hubble parameter is known as the Hubble constant 
and is denoted by $H_0$. Defining the redshift $z = a_0/a - 1 $, and 
the present value of the density parameters,
\begin{equation}
\label{eq:density parameters}
   \Omega_{m0} = \frac{8\pi G\rho_0}{3H_0^2}, \ \
   \Omega_{K0}=\frac{-K^2}{(H_0 a_0)^2}, \ \ 
   \Omega_\Lambda = \frac{\Lambda}{3H_0^2},
\end{equation}
in the $\Lambda$CDM model we can rewrite Eq.\ (\ref{eq:F1}) as
\begin{equation}
\label{eq:LCDMFM}
H^2(z; H_0, \textbf{p}) = H_0^2 \left[\Omega_{m0} (1+z)^3 + \Omega_{\Lambda}
                          + (1-\Omega_{m0}-\Omega_\Lambda) (1+z)^2 \right],
\end{equation}
where we have made use of $\Omega_{K0} = 1-\Omega_{m0}-\Omega_{\Lambda}$.
This is the Friedmann equation of the $\Lambda$CDM model with spatial 
curvature. In this model the cosmological parameters 
$\textbf{p}=({\Omega_{m0},\Omega_\Lambda})$. Here $\Omega_{m0}$ is 
the non-relativistic (baryonic and cold dark) matter energy density 
parameter at the present time. Below we shall have need for the
dimensionless Hubble parameter $E(z) = H(z)/H_0$. 

It has become popular to parametrize time-varying dark energy as a spatially 
homogeneous $X$-fluid, with a constant equation of state parameter 
$\omega_{\rm X} = p_{\rm X}/\rho_{\rm X} <-1/3 $. With this XCDM
parametrization the Friedmann equation takes the form
\begin{equation}
   H^2(z; H_0, \textbf{p}) = H_0^2 [\Omega_{m0}(1+z)^3 + 
      (1 - \Omega_{m0}) (1+z)^{3(1+\omega_{\rm X})}], 
\end{equation}
where for computational simplicity we consider only flat spatial 
hypersurfaces, and the model parameters $\textbf{p}= (\Omega_{m0} ,
\omega_{\rm X})$. The XCDM parametrization is incomplete, as it
cannot describe the evolution of energy density inhomogeneities.

The simplest complete and consistent dynamical dark energy model is 
$\phi$CDM. In this model dark energy is a slowly-rolling scalar field 
$\phi$ with an, e.g., inverse-power-law potential energy density 
$V(\phi)=\kappa m_p^2 \phi^{-\alpha} $ where $m_p=1/\sqrt{G}$ is 
the Planck mass and $\alpha$ is a non-negative free parameter that 
determines $\kappa$. The scalar field part of the $\phi$CDM model action is
\begin{equation}
   S=\frac{m_p^2}{16\pi}\int{\sqrt{-g}\left( \frac{1}{2} ~g^{\mu \nu}
   \partial_\mu \phi \partial_\nu \phi - \kappa m_p^2 \phi^{-\alpha} 
   \right) d^4x},
\end{equation}
with corresponding scalar field equation of motion 
\begin{equation}
\label{eq:dotphi} 
    \ddot{\phi} + 3 \frac{\dot{a}}{a}\dot{\phi} -
    \kappa \alpha m_p^2 \phi^{-(\alpha+1)} = 0.
\end{equation}
In the spatially-flat case the Friedmann equation is
\begin{equation}
\label{eq:phicdmfriedman}
   H^2(z; H_0, \textbf{p}) = \frac{8\pi G}{3}(\rho_m +\rho_\phi)
   = H_0^2[\Omega_{m0}(1+z)^3+\Omega_\phi (z,\alpha)],
\end{equation}
with scalar field energy density given by
\begin{equation}
\label{eq:rhom} 
   \rho_\phi = \frac{m_p^2}{16\pi} \left({\frac{1}{2}}\dot{\phi}^2 
   + \kappa m_p^2 \phi^{-\alpha}\right).
\end{equation}
Solving the coupled differential 
Eqs.\ ({\ref{eq:dotphi}})---({\ref{eq:rhom}}), 
with the initial conditions described in 
\cite{Peebles&Ratra1988}, allows for a numerical computation of 
the Hubble parameter $H(z)$. In this case the model parameter 
set is $\textbf{p}=(\Omega_{m0},\alpha)$.

\section{Constraints from the $H(z)$ data}
\label{HzData}

We use 21 independent $H(z)$ data points \citep{simon05, gaztanaga09, Stern2010, 
moresco12}, listed in Table 1, to constrain cosmological model
parameters. The observational data consist of measurements of the 
Hubble parameter $H_{\rm obs}(z_i)$ at redshifts $z_i$, with the 
corresponding one standard deviation uncertainties $\sigma_i$.

To constrain cosmological parameters $\textbf{p}$ of the 
models of interest we compute the $\chi_{H}^2$ function 
\begin{equation}
\label{eq:chi2H} 
\chi_{H}^2 (H_0, \textbf{p}) =
\sum_{i=1}^{21}\frac{[H_{\rm th} (z_i; H_0, \textbf{p})-H_{\rm
obs}(z_i)]^2}{\sigma^2_i}.
\end{equation}
where $H_{\rm th} (z_i; H_0, \textbf{p})$ is the model-predicted
value of the Hubble parameter. As discussed in Sec.\ {\ref{equations}}, 
$ H_{\rm th} (z_i; H_0, \textbf{p})=H_0 E(z;\textbf{p})$, so
from Eq. ({\ref{eq:chi2H}}) we find
\begin{equation}
\label{eq:chi2Hs} 
\chi_{H}^2 (H_0, \textbf{p}) =
H_0^2 \sum_{i=1}^{21}\frac{E^2(z_i; \textbf{p})}{\sigma^2_i}
-2H_0 \sum_{i=1}^{21}\frac{H_{\rm obs}(z_i)E(z_i; \textbf{p})}{\sigma^2_i}
+\sum_{i=1}^{21}\frac{H^2_{\rm obs}(z_i)}{\sigma^2_i}.
\end{equation}

$\chi_{H}^2$ depends on the model parameters $\textbf{p}$ as well as 
on the nuisance parameter $H_0$ whose value is not known exactly.
We assume that the distribution of $H_0$ is a Gaussian with one 
standard deviation width  $\sigma_{H_0}$ and mean $\bar{H_0}$. We
can then build the posterior likelihood function $\mathcal{L}_{H}(\textbf{p})$
that depends only on the $\textbf{p}$ by integrating the product 
of exp$(-\chi_H^2 /2)$ and the $H_0$ prior likelihood function 
exp$[-(H_0-\bar H_0)^2/(2\sigma^2_{H_0})]$ \citep[see, e.g.,][]{Ganga1997}, 
\begin{equation}
\label{eq:likely}
\mathcal{L}_{H}(\textbf{p})=\frac{1}{\sqrt{2\pi \sigma^2_{H_0}}}
   \int \limits_0^\infty e^{-\chi_H^2(H_0,\textbf{p})/2} 
   e^{-(H_0-\bar H_0)^2/(2\sigma^2_{H_0})} dH_0.
\end{equation}
Defining
\begin{equation}
\alpha=\frac{1}{\sigma_{H_0}^2}+\sum_{i=1}^{21}\frac{E^2(z_i; \textbf{p})}{\sigma^2_i},~
\beta=\frac{\bar H_0}{\sigma_{H_0}^2}+\sum_{i=1}^{21}\frac{H_{\rm obs}(z_i)E(z_i; \textbf{p})}{\sigma^2_i},~
\gamma=\frac{\bar H_0^2}{\sigma_{H_0}^2}+\sum_{i=1}^{21}\frac{H^2_{\rm obs}(z_i)}{\sigma^2_i}, 
\end{equation}
the integral can be expressed in terms of the error function,\footnote{$
{\rm erf}(x)={\frac{2}{\sqrt{\pi}}}\int \limits_{0}^{x}{{e^{-t^2}dt}}$.}
\begin{equation}
   \mathcal{L}_H(\textbf{p})=\frac{1}{2 \sqrt{\alpha~ \sigma_{H_0}^2}} 
   \exp \left[-\frac{1}{2}\left({\gamma}-\frac{\beta^2}{\alpha}\right)\right]
   \left[ 1 + \mathrm{erf}\left({\frac{\beta }{\sqrt{2\alpha}}}\right)\right].
\end{equation}

\begin{figure}[t]
\centering
  \includegraphics[width=80mm]{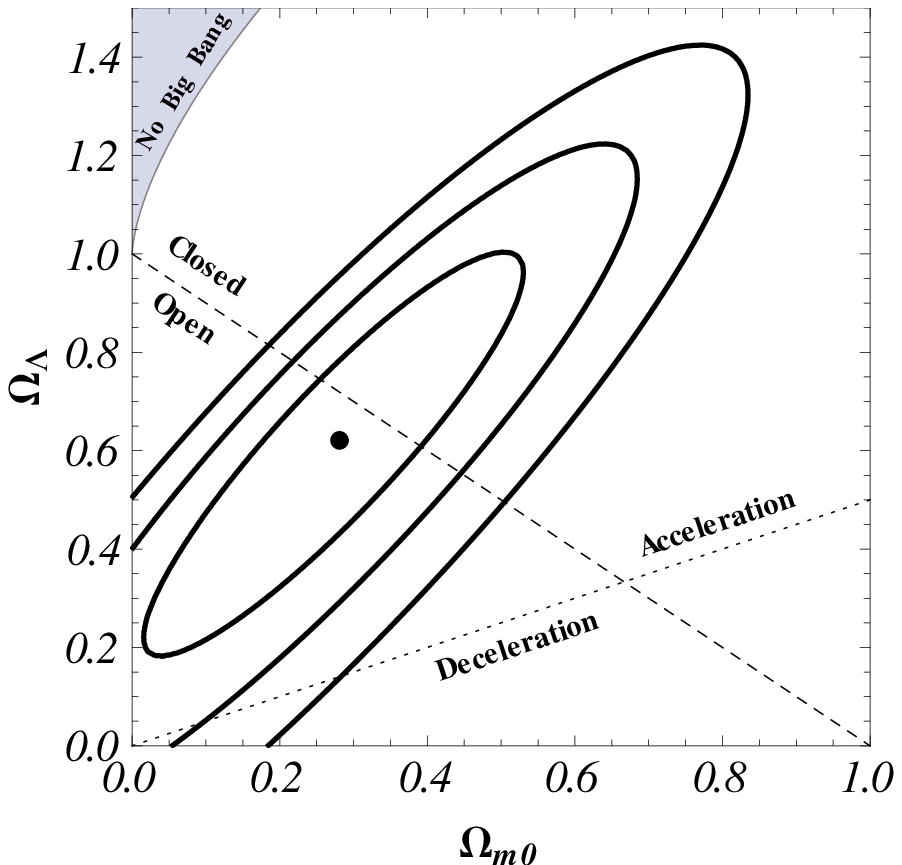}
  \includegraphics[width=80mm]{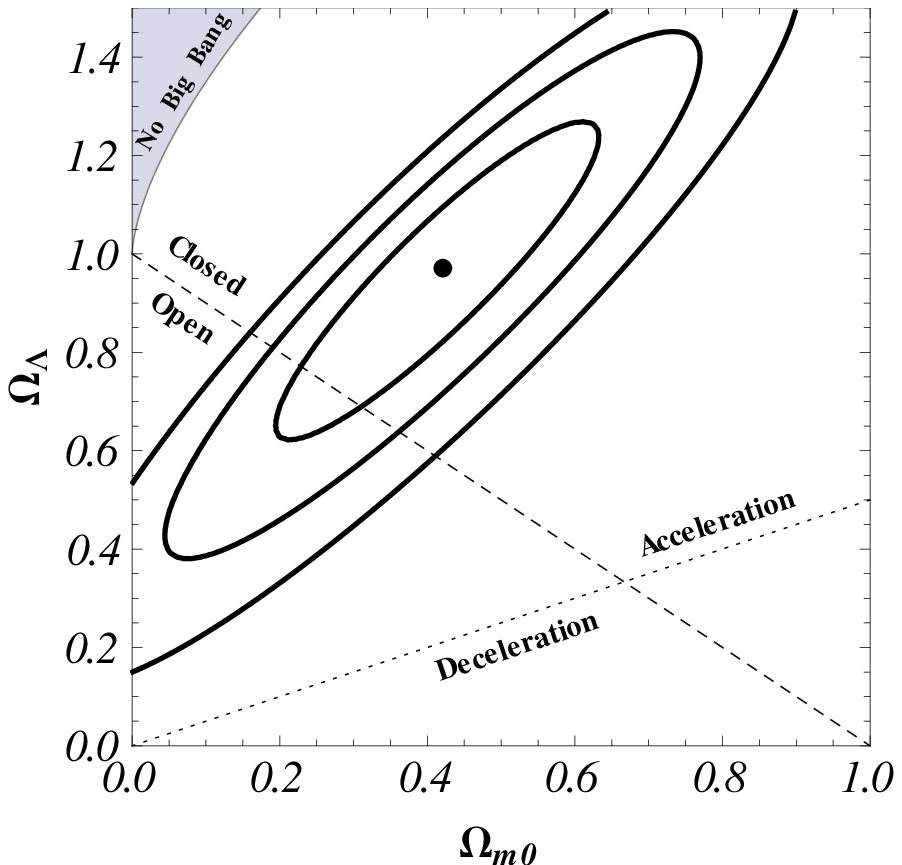}

\caption{
Solid lines shows 1, 2, and 3 $\sigma$ constraint contours for 
the $\Lambda$CDM model from the $H(z)$ data. The left panel is for the 
$H_0 = 68 \pm 2.8$ km s$^{-1}$ Mpc$^{-1}$ prior and the right 
panel is for the $H_0 = 73.8 \pm 2.4$ km s$^{-1}$ Mpc$^{-1}$ one.
Thin dot-dashed lines in the left panel are 1, 2, and 3 $\sigma$ 
contours reproduced from \cite{Chen2011b}, where the prior is 
$H_0 = 68 \pm 3.5$ km s$^{-1}$ Mpc$^{-1}$; the empty circle is the 
corresponding  best-fit point.
The dashed diagonal lines correspond to spatially-flat models, the 
dotted lines demarcate zero-acceleration models, and the shaded area 
in the upper left-hand corners are the region  for which there is no 
big bang. The filled black circles correspond to best-fit points. For
quantitative details see Table \ref{tab:results-1}.
} \label{fig:LCDM_Hz}
\end{figure}


\begin{figure}[t]
\centering
  \includegraphics[width=80mm]{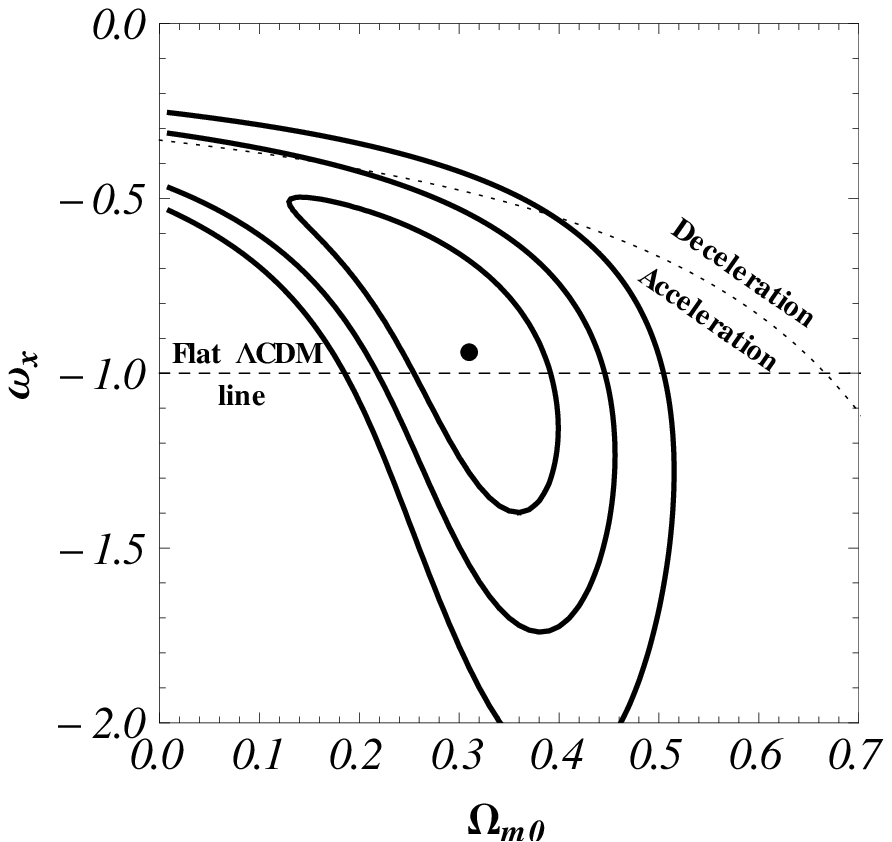}
  \includegraphics[width=80mm]{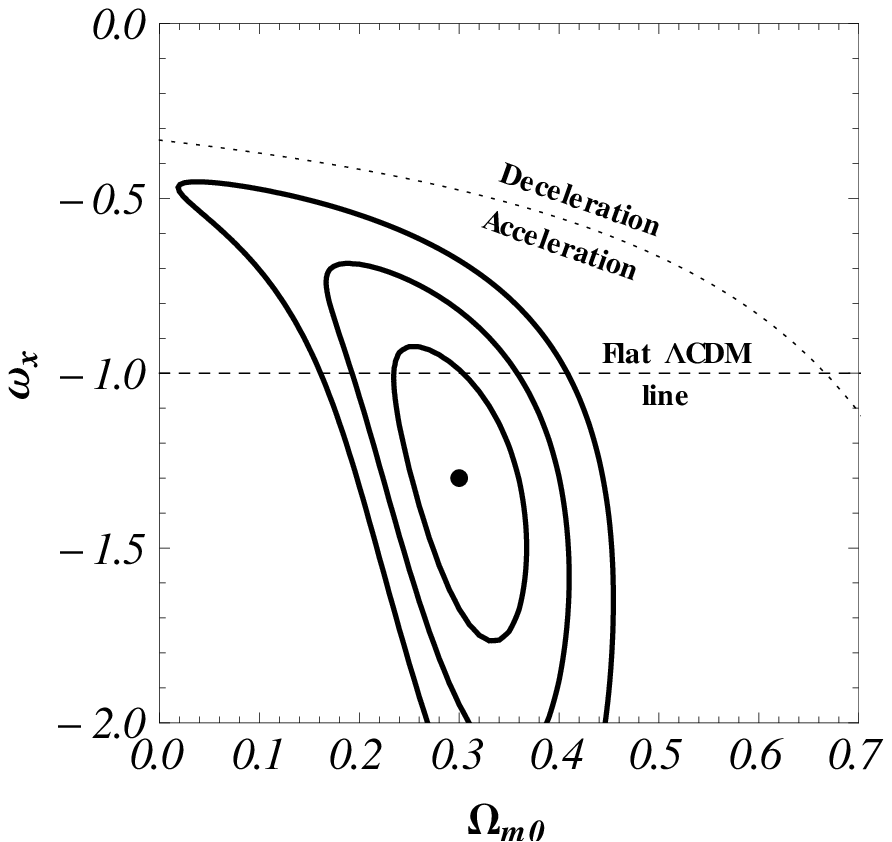}

\caption{
Solid lines shows 1, 2, and 3 $\sigma$ constraint contours for the XCDM
parametrization from the $H(z)$ data. The left panel is for the 
$H_0 = 68 \pm 2.8$ km s$^{-1}$ Mpc$^{-1}$ prior and the right 
panel is for the $H_0 = 73.8 \pm 2.4$ km s$^{-1}$ Mpc$^{-1}$ one.
Thin dot-dashed lines in the left panel are 1, 2, and 3 $\sigma$ 
contours reproduced from \cite{Chen2011b}, where the prior is 
$H_0 = 68 \pm 3.5$ km s$^{-1}$ Mpc$^{-1}$; the empty circle is the 
corresponding best-fit point.
The dashed horizontal lines at $\omega_{\rm X} = -1$ correspond to 
spatially-flat $\Lambda$CDM models and the curved dotted lines demarcate 
zero-acceleration models. The filled black circles correspond to best-fit
points. For quantitative details see Table \ref{tab:results-1}.
} \label{fig:XCDM_Hz}
\end{figure}

\begin{figure}[t]
\centering
  \includegraphics[width=80mm]{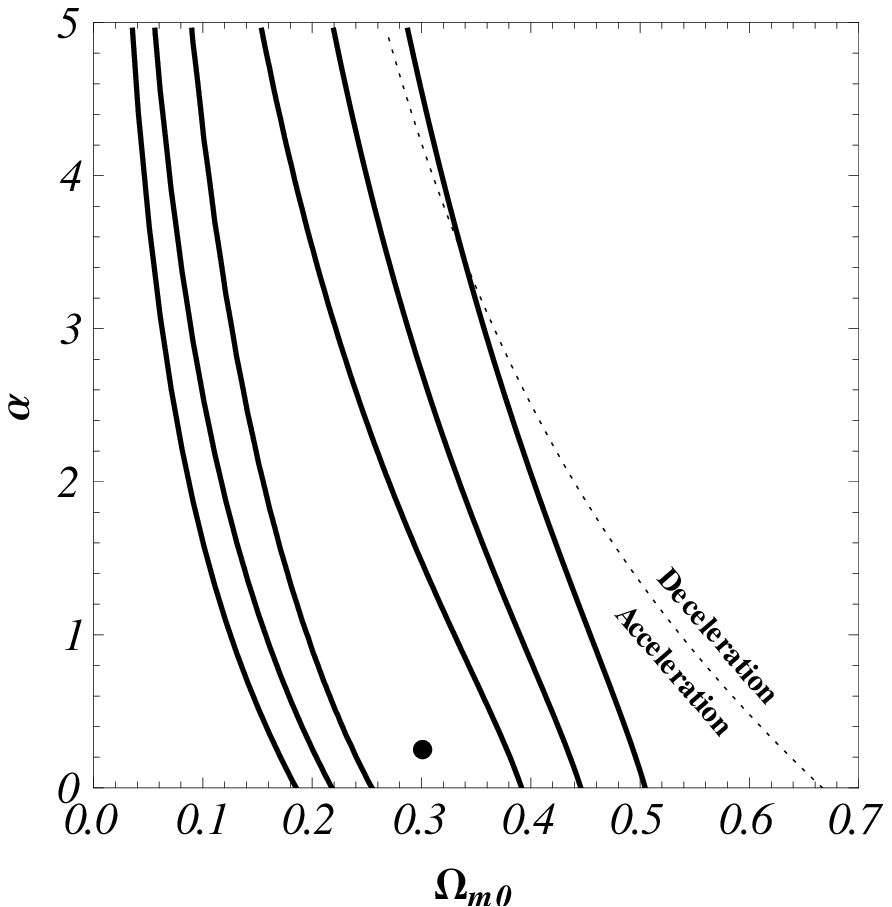}
  \includegraphics[width=80mm]{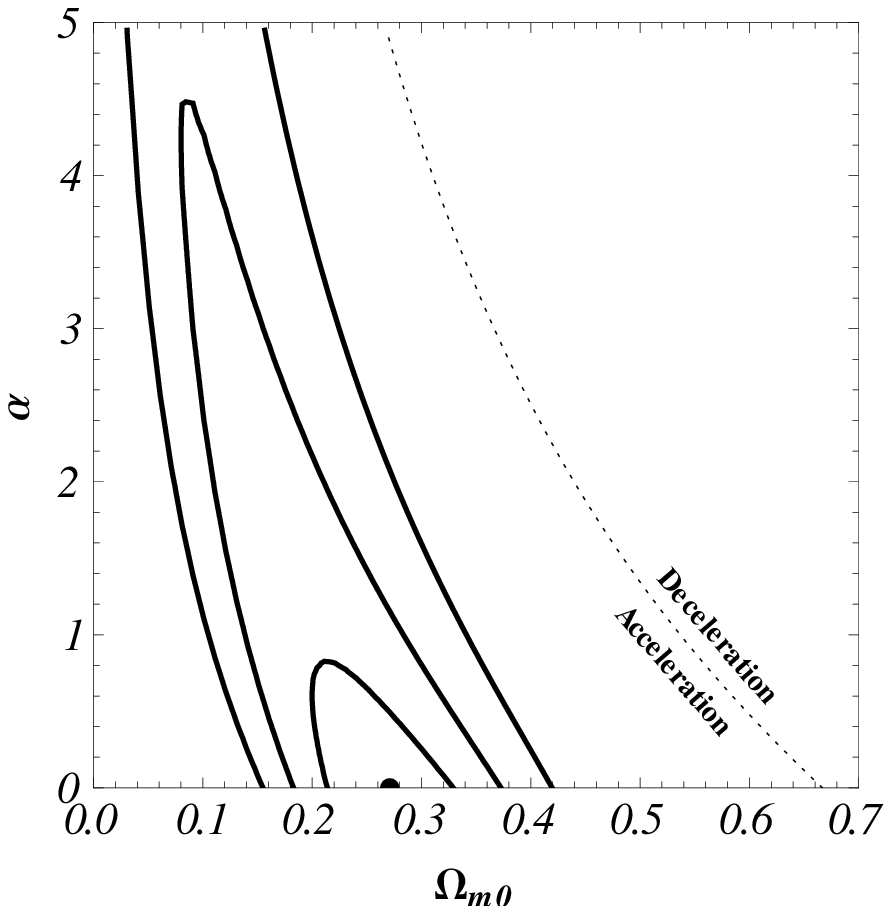}

\caption{
Solid lines shows 1, 2, and 3 $\sigma$ constraint contours for 
the $\phi$CDM model from the $H(z)$ data. The left panel is for the 
$H_0 = 68 \pm 2.8$ km s$^{-1}$ Mpc$^{-1}$ prior and the right 
panel is for the $H_0 = 73.8 \pm 2.4$ km s$^{-1}$ Mpc$^{-1}$ one.
Thin dot-dashed lines in the left panel are 1, 2, and 3 $\sigma$ 
contours reproduced from \cite{Chen2011b}, where the prior is 
$H_0 = 68 \pm 3.5$ km s$^{-1}$ Mpc$^{-1}$; the empty circle is 
the corresponding best-fit point.
The horizontal axes at $\alpha = 0$ correspond to spatially-flat 
$\Lambda$CDM models and the curved dotted lines demarcate 
zero-acceleration models. The filled black circles correspond to best-fit
points. For quantitative details see Table \ref{tab:results-1}.
} \label{fig:phiCDM_Hz}
\end{figure}

We maximize the likelihood $\mathcal{L}_H(\textbf{p})$, 
or equivalently minimize $\chi_H^2(\textbf{p}) = -2 
\mathrm{ln}\mathcal{L}_{H}(\textbf{p})$, with respect to the 
parameters $\textbf{p}$ to find the best-fit parameter values 
$\mathbf{p_0}$. In the models we consider $\chi_H^2$ depends on 
two parameters. We define 1$\sigma$, 2$\sigma$, and 3$\sigma$ 
confidence intervals as two-dimensional parameter sets bounded by 
$\chi_H^2(\textbf{p}) = \chi_H^2(\mathbf{p_0})+2.3,~\chi_H^2(\textbf{p}) = 
\chi_H^2(\mathbf{p_0})+6.17$, and $\chi_H^2(\textbf{p}) = 
\chi_H^2(\mathbf{p_0})+11.8$, respectively.

Even though the precision of measurements of the Hubble constant have
greatly improved over the last decade, the concomitant improvement 
in the precision of other cosmological measurements means that in some
cases the Hubble constant uncertainty still significantly affects 
cosmological parameter estimation. For a recent example see 
\citet{calabrese12}. The values of $\bar{H_0}\pm\sigma_{H_0}$ 
that we use in this paper are 68 $\pm$ 2.8 km s$^{-1}$ Mpc$^{-1}$ and 
73.8 $\pm$ 2.4 km s$^{-1}$ Mpc$^{-1}$. The first is from a median 
statistics analysis \citep{Gott2001} of 553 measurements of $H_0$
\citep{Chen2011a}; this estimate has been remarkably stable for over 
a decade now \citep{Gott2001, Chen2003}. The second value is the most
precise recent one, based on HST measurements \citep{Riess2011}. Other 
recent measurements are not inconsistent with at least one of the two 
values we use as a prior \citep[see, e.g.,][]{Freedman2012, Sorce2012, 
Tammann2012}. 

\begin{table}[htb]
\begin{center}
\begin{tabular}{ccccccc} 
\hline\hline 
{} &  
\multicolumn{2}{c}{$H(z)$}
&\multicolumn{2}{c}{SNeIa}
&\multicolumn{2}{c}{BAO}\\
Model and prior & $\chi^2_{\rm min}$ & B.F.P & $\chi^2_{\rm min}$ & B.F.P & $\chi^2_{\rm min}$ & B.F.P\\ 
\hline \hline 
$\Lambda$CDM & \multirow{2}{*}{$14.6$} & $\Omega_{m0}=0.28$ & \multirow{4}{*}{545} & \multirow{2}{*}{$\Omega_{m0}=0.29$} & \multirow{4}{*}{5.5} & \multirow{2}{*}{$\Omega_{m0}=0.27$} \\ 
$h = 0.68 \pm 0.028$ &  & $\Omega_{\Lambda}=0.62$ &  & {} &  & {} \\ 
\cline{1-3}

$\Lambda$CDM & \multirow{2}{*}{$14.6$} & $\Omega_{m0}=0.42$ & \multirow{4}{*}{} & \multirow{2}{*}{$\Omega_{\Lambda}$=0.69} & \multirow{2}{*}{} & \multirow{2}{*}{$\Omega_{\Lambda}$=0.87} \\ 
$h = 0.738 \pm 0.024$ &  & $\Omega_{\Lambda}=0.97$ &  & \multirow{4}{*}{} &  & \multirow{4}{*}{}\\ 
\hline
\hline

XCDM & \multirow{2}{*}{$14.6$} & $\Omega_{m0}=0.31$ & \multirow{4}{*}{545} & \multirow{2}{*}{$\Omega_{m0}=0.29$} & \multirow{4}{*}{$5.5$} & \multirow{2}{*}{$\Omega_{m0}=0.27$} \\ 
$h = 0.68 \pm 0.028$ &  & $\omega_{X}=-0.94$ &  & {} &  & {} \\
\cline{1-3}

XCDM & \multirow{2}{*}{$14.6$} & $\Omega_{m0}=0.30$ & \multirow{4}{*}{} & \multirow{2}{*}{$\omega_{X}=-0.99$} & \multirow{2}{*}{} & \multirow{2}{*}{$\omega_{X}=-1.21$} \\ 
$h = 0.738 \pm 0.024$ &  & $\omega_{X}=-1.3$ &  & \multirow{4}{*}{} &  & \multirow{4}{*}{}\\ 
\hline 
\hline

$\phi$CDM & \multirow{2}{*}{$14.6$} & $\Omega_{m0}=0.30$ & \multirow{4}{*}{545} & \multirow{2}{*}{$\Omega_{m0}=0.27$} & \multirow{4}{*}{5.9} & \multirow{2}{*}{$\Omega_{m0}=0.30$} \\ 
$h = 0.68 \pm 0.028$ &  & $\alpha=0.25$ &  & {} &  & {} \\
\cline{1-3}

$\phi$CDM & \multirow{2}{*}{$15.6$} & $\Omega_{m0}=0.27$ & \multirow{4}{*}{} & \multirow{2}{*}{$\alpha =0.20$} & \multirow{2}{*}{} & \multirow{2}{*}{$\alpha =0.00$} \\ 
$h = 0.738 \pm 0.024$ &  & $\alpha =0.00$ &  & \multirow{4}{*}{} &  & \multirow{4}{*}{}\\ 
\hline
\hline  
\end{tabular}

\caption{The minimum value of $\chi^2$ and the corresponding best-fit 
points (B.F.P) which maximize the likelihood for the three individual
data sets. The SNIa values are for the case including systematic errors. 
Ignoring SNIa systematic errors, for the $\Lambda$CDM model
$\chi_{SN}^2(\mathbf{p_0})=562$, at $(\Omega_{m0}, \Omega_\Lambda) = (0.28,0.73)$; for 
the XCDM case $\chi_{SN}^2(\mathbf{p_0})=562$ at $(\Omega_{m0}, 
\omega_{\rm X}) = (0.28,-1.01)$; and for the $\phi$CDM model 
$\chi_{SN}^2(\mathbf{p_0})=562$, at $(\Omega_{m0}, \alpha) = (0.27,0.05)$.}
\label{tab:results-1}
\end{center}
\end{table}

Figures \ref{fig:LCDM_Hz}---\ref{fig:phiCDM_Hz} show the constraints
from the $H(z)$ data for the three dark energy models we consider, and
for the two different $H_0$ priors. Table \ref{tab:results-1} lists 
the best fit parameter values. Comparing these plots with Figs.\ 
1---3 of \cite{Chen2011b}, whose 1, 2 and 3 $\sigma$ constraint 
contours are reproduced here as dot-dashed lines in the left panels of 
Figs \ref{fig:LCDM_Hz}---\ref{fig:phiCDM_Hz}, we see that the 
contours derived from the new data are more constraining, 
by about a standard deviation, because of the 8 new, more
precise, \citet{moresco12} data points used here. On comparing the 
left and right panels in these three figures, we see that the constraint
contours are quite sensitive to the value of $H_0$ used, as well as to 
the uncertainty associated with the Hubble constant measurement.

\section{Constraints from the SNIa data}
\label{SNeIa}

While the $H(z)$ data provide tight constraints on a linear
combination of cosmological parameters, the very elongated constraint
contours of Figs.\ \ref{fig:LCDM_Hz}---\ref{fig:phiCDM_Hz}
imply that these data alone cannot significantly discriminate
between cosmological models. To tighten the constraints we must add other 
data to the mix.

The second set of data that we use are the Type Ia supernova data 
from the \cite{suzuki12}
Union2.1 compilation of 580 SNIa distance modulus $\mu_{\rm obs}(z_i)$ 
measurements at measured redshifts $z_i$ (covering the redshift range
of 0.015 to 1.414) with associated one standard deviation  
uncertainties $\sigma_i$. The predicted distance modulus is 
\begin{equation}
\label{eq:distance modulus theoratical}
\mu_{\rm th} (z_i; H_0, \textbf{p}) = \underbrace{5 ~ \mathrm{log}_{10}\left(3000 ~ y(z) (1+z)\right) +25}_{=\mu_{0}} ~ -5 ~ \mathrm{log}_{10}(h),
\end{equation} 
where $H_0 = 100 h$ km s$^{-1}$ Mpc$^{-1}$ and 
$y(z)$ is the dimensionless coordinate distance,
\begin{equation}
\label{eq:Angular size distance}
y(z) = \left\{
     \begin{array}{lr}
       {\frac{a_0 H_0 }{K}}~\mathrm{sin}\left(\frac{K}{a_0H_0} 
       \int \limits_{0}^{z}{\frac{dz'}{E(z')}}\right) & K^2 > 0\\
       \int \limits_{0}^{z}{\frac{dz'}{E(z')}} & K^2 = 0\\
       {\frac{a_0 H_0}{\sqrt{-K^2}}} ~\mathrm{sinh}\left(\frac{\sqrt{-K^2}}
       {a_0H_0} \int \limits_{0}^{z}{\frac{dz'}{E(z')}}\right) &  K^2 < 0.\\
     \end{array}
   \right.
\end{equation}

As the SNIa distance modulus measurements $\mu_{\rm obs}$ are correlated, 
$\chi^2$ is defined as 
\begin{equation}
\label{eq:chiSN-1}
\chi_{SN}^{2}(h,\textbf{p})=\Delta\boldsymbol\mu^T~{\mathcal{C}}^{-1}~
\Delta\boldsymbol{\mu}.
\end{equation}
Here $\Delta \boldsymbol\mu $ is a vector of differences 
$\Delta{\mu_i}= \mu_{\rm th}(z_i;H_0,\textbf{p}) -\mu_{\rm obs}(z_i)$, 
and $\mathcal{C}^{-1}$ is the inverse of the 580 by 580 Union 2.1 compilation 
covariance matrix. In index notation, 
\begin{equation}
\label{eq:chiSN-2}
\chi^{2}_{SN}(h,\textbf{p})=\sum_{\alpha, \beta}\left[\mu_0 -5 
    \mathrm{log}_{10} h - \mu_{\rm obs} \right]_\alpha
    (\mathcal{C}^{-1})_{\alpha \beta}
    \left[\mu_0 - 5 \mathrm{log}_{10} h - \mu_{\rm obs}
    \right]_\beta.
\end{equation}
The covariance matrix is symmetric so this can be written as
\begin{equation}
\label{eq:chiSN-3}
\chi^{2}_{SN}(h,\textbf{p})=A(\textbf{p}) - 
10 B(\textbf{p}) \mathrm{log}_{10}(h)+ 25 C [\mathrm{log}_{10}(h)]^2
\end{equation}
where
\begin{equation}
\label{eq:chiSN-3a}
\begin{array}{lr}
A(\textbf{p})= \sum\limits_{\alpha, \beta} (\mu_{0}-\mu_{\rm obs})_{\alpha}\ 
    (\mathcal{C}^{-1})_{\alpha \beta}\ (\mu_{0}-\mu_{\rm obs})_{\beta}\\
B(\textbf{p})= \sum\limits_\alpha (\mu_{0}-\mu_{\rm obs})_\alpha\sum\limits_\beta 
    (\mathcal{C}^{-1})_{\alpha \beta} \\
C= \sum\limits_{\alpha,\beta} (\mathcal{C}^{-1})_{\alpha \beta}.
\end{array}
\end{equation}

The corresponding likelihood function, when considering a flat $H_0$ prior,
is
\begin{equation}
 \mathcal{L}_{SN}(\textbf{p}) =  \int\limits_{0}^{\infty}
     {e^{-\chi^2_{SN}(h,\textbf{p})/2} dh}.     
\end{equation}
 Defining 
\begin{equation}
\delta = \frac{25C}{2\mathrm{(ln10)^2}}~, 
 ~~~\varepsilon = \frac{B(\textbf{p})\mathrm{ln10}}{5C},
 \nonumber
\end{equation}
the above integral takes the form
\begin{eqnarray}
\label{eq:chiSN-4}
 \mathcal{L}_{SN}(\textbf{p}) =  \sqrt{\frac{\pi}{\delta}} 
{\rm exp}\left[-\frac{1}{2}\left(A(\textbf{p})-\frac{B^2(\textbf{p})}{C}
-2\varepsilon - \frac{1}{2\delta^2}\right)\right].
\end{eqnarray}
The $h$-independent 
\begin{equation}
\label{eq:chiSN-5}
\chi^2_{SN}(\textbf{p}) =  -2~\mathrm{ln} \mathcal{L}_{SN}(\textbf{p}) 
 = A(\textbf{p})-\frac{B^2(\textbf{p})}{C} - 
   \frac{2 \mathrm{ln}(10)}{5C}B(\textbf{p})- Q,
\end{equation}
where $Q$ is a constant that does not depend on the model parameters $\textbf{p}$,
\begin{equation}
\label{eq:G}
Q=\frac{2(\mathrm{ln}10)^4}{625~C^2}+2~\mathrm{ln}\left(\frac{2\pi(\mathrm{ln}10)^2}{25~C}\right), \nonumber
\end{equation}
and so can be ignored.
We minimize $\chi_{SN}^2(\textbf{p})$ with respect to the model parameters 
$\textbf{p}$ to find the best-fit parameter values $\mathbf{p_0}$ and
constraint contours.

\begin{figure}[p]
\centering
  \includegraphics[width=79.0mm]{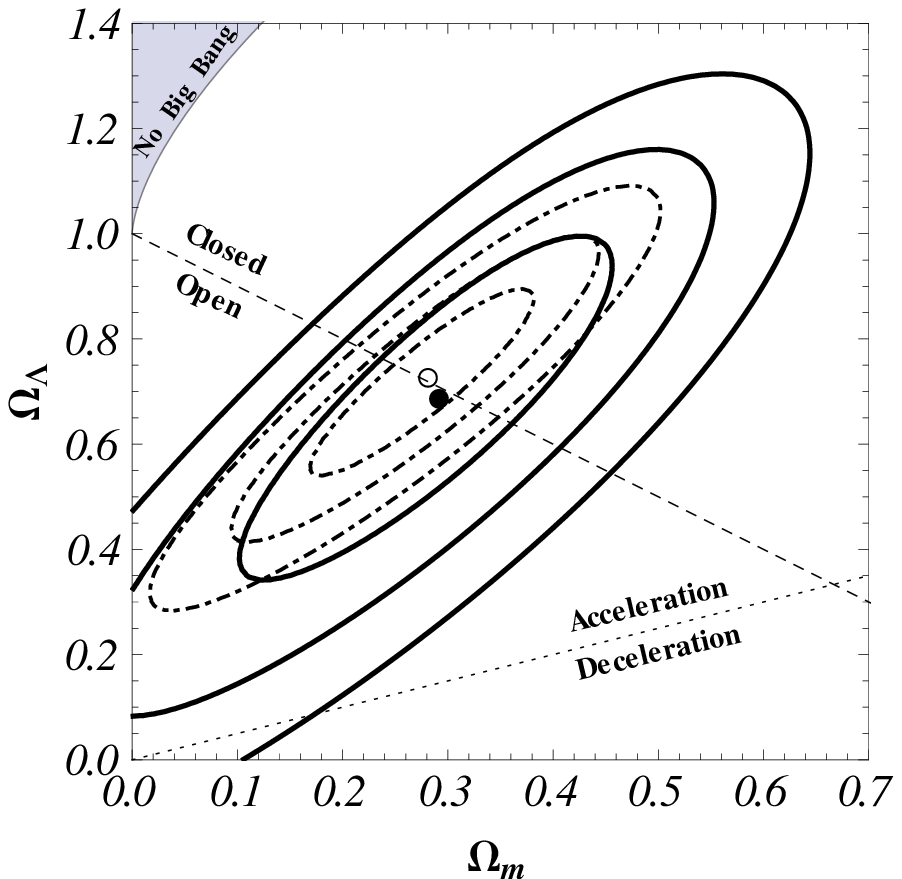}
  \includegraphics[width=82.0mm]{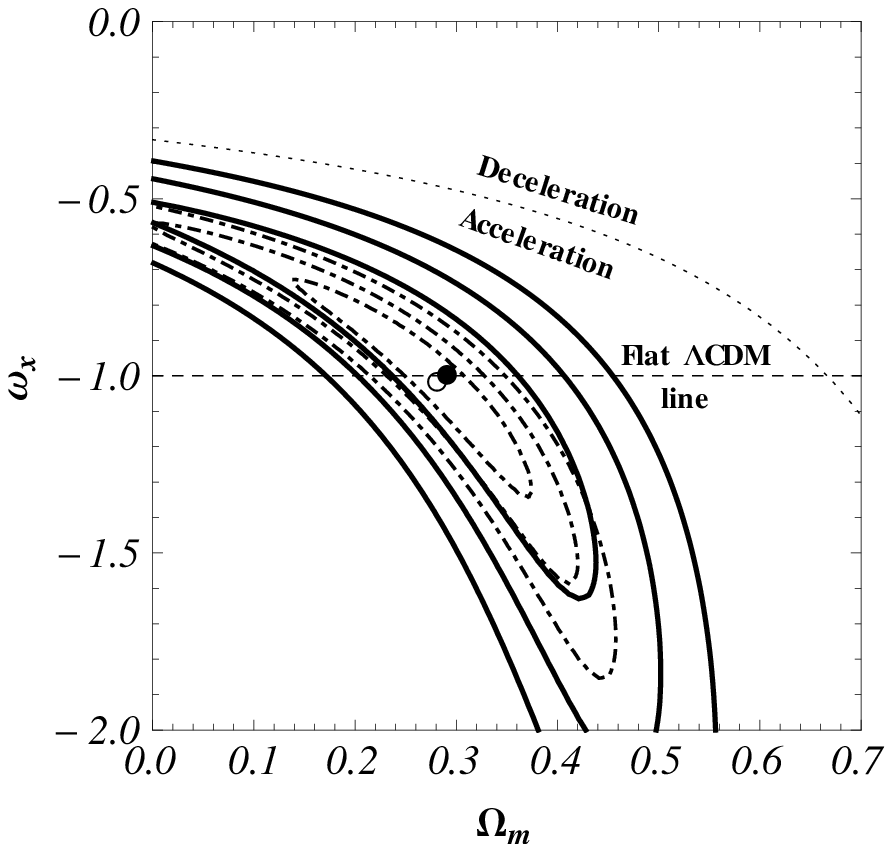}
  \includegraphics[width=79.mm]{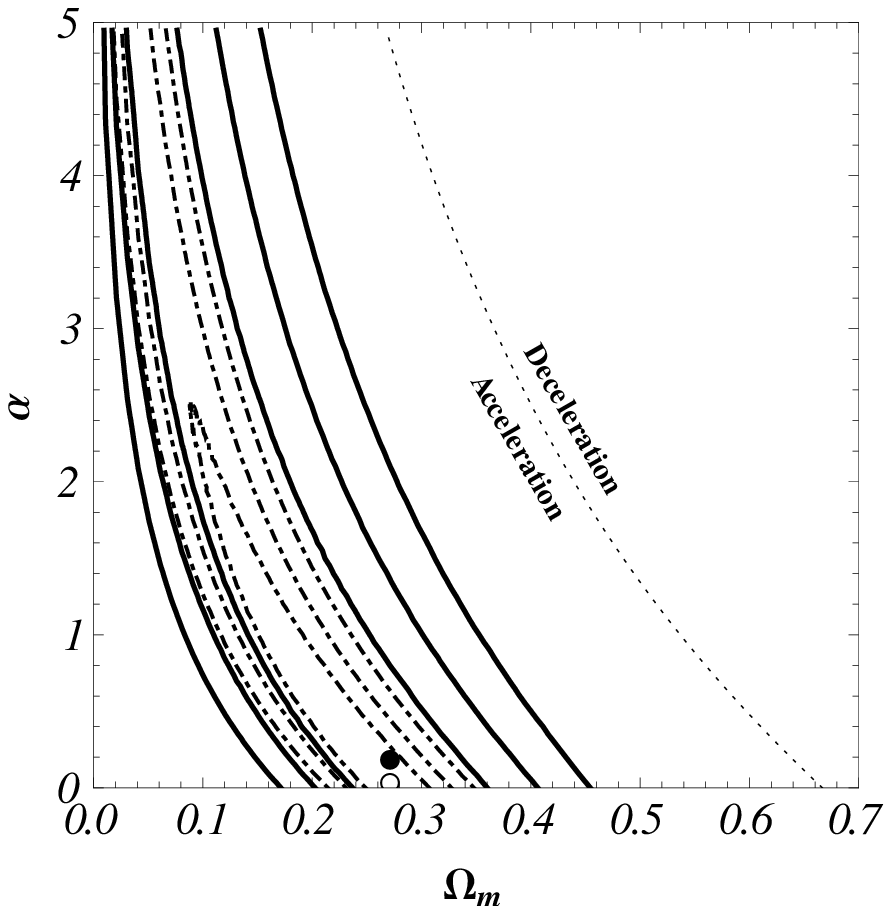}
\caption{
Thick solid (dot-dashed) lines are 1, 2, and 3$-\sigma$ constraint contours 
from SNIa data with (without) systematic errors. Filled (open) circles 
demarcate likelihood maxima for the case of data with (without) systematic 
errors. The top left plot is for the $\Lambda$CDM model, the top right 
plot is for the XCDM parametrization, and the bottom one is for the 
$\phi$CDM model. For quantitative details see Table \ref{tab:results-1}.
} \label{fig:SNEIA-lxCDM}
\end{figure}

Figure \ref{fig:SNEIA-lxCDM} shows constraints from the SNIa data on 
the three dark energy models we consider here. For the $\Lambda$CDM 
model and the XCDM parametrization the constraints shown in Fig.\ 
\ref{fig:SNEIA-lxCDM} are in very good agreement with those in Figs.\ 
5 and 6 of \cite{suzuki12}. The $\phi$CDM model SNIa data constraints 
shown in Fig.\ \ref{fig:SNEIA-lxCDM} have not previously been computed. 
Comparing the SNIa 
constraints of Fig.\ \ref{fig:SNEIA-lxCDM} to those which follow 
from the $H(z)$ data, Figs.\ \ref{fig:LCDM_Hz}---\ref{fig:phiCDM_Hz},
it is clear that SNIa data provide tighter constraints on the  
$\Lambda$CDM model. For the XCDM case both SNIa data and $H(z)$ data
provide approximately similar constraints, while the SNIa constraints
are somewhat more restrictive than the $H(z)$ ones for the $\phi$CDM
model. However, in general, the SNIa constraints are not very significantly 
more restrictive than the $H(z)$ constraints, which is a remarkable
result. It is also reassuring that both data favor approximately 
similar regions of parameters space, for all three models we consider.
However, given that the degeneracy in parameter space is similar for the
$H(z)$ and SNIa data, a joint analysis of just these two data sets is 
unlikely to greatly improve the constraints.

\section{Constraints from the BAO data}
\label{BAO}

In an attempt to further tighten the cosmological parameter constraints,
we now include BAO data in the analysis. To constrain cosmological 
parameters using 
BAO data we follow the procedure of \cite{blake11}. To derive the BAO
constraints we make use of the distance parameter $D_V (z)$, a 
combination of the angular diameter distance and the Hubble parameter, 
given by 
\begin{equation}
\label{eq:D_V}
D_V(z)= \left[(1 + z)^2 d_A(z)^2 \frac{c~z}{H(z)}\right]^{1/3}.
\end{equation}
Here $d_A(z)$ is the angular diameter distance
\begin{equation}
\label{eq:D_A}
d_A(z)=\frac{y(z)}{H_0(1+z)}
\end{equation}
where $y(z)$ is the dimensionless coordinate distance given in Eq.\ 
({\ref{eq:Angular size distance}}).

\begin{figure}[p]
\centering
  \includegraphics[width=79mm]{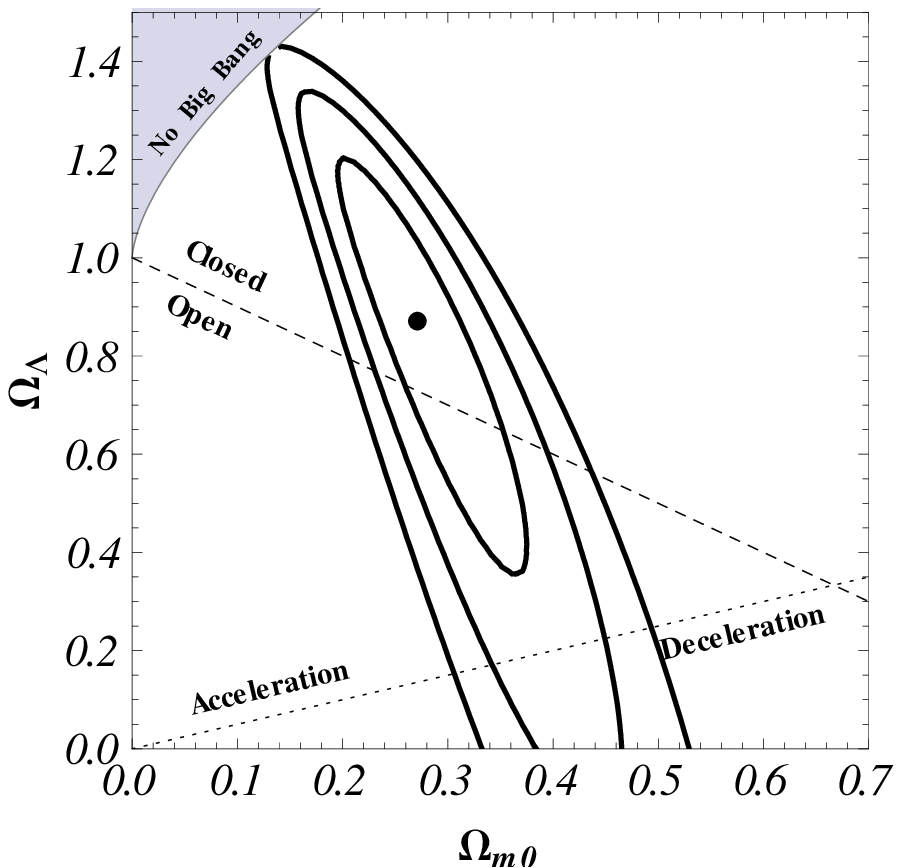}
  \includegraphics[width=82.5mm]{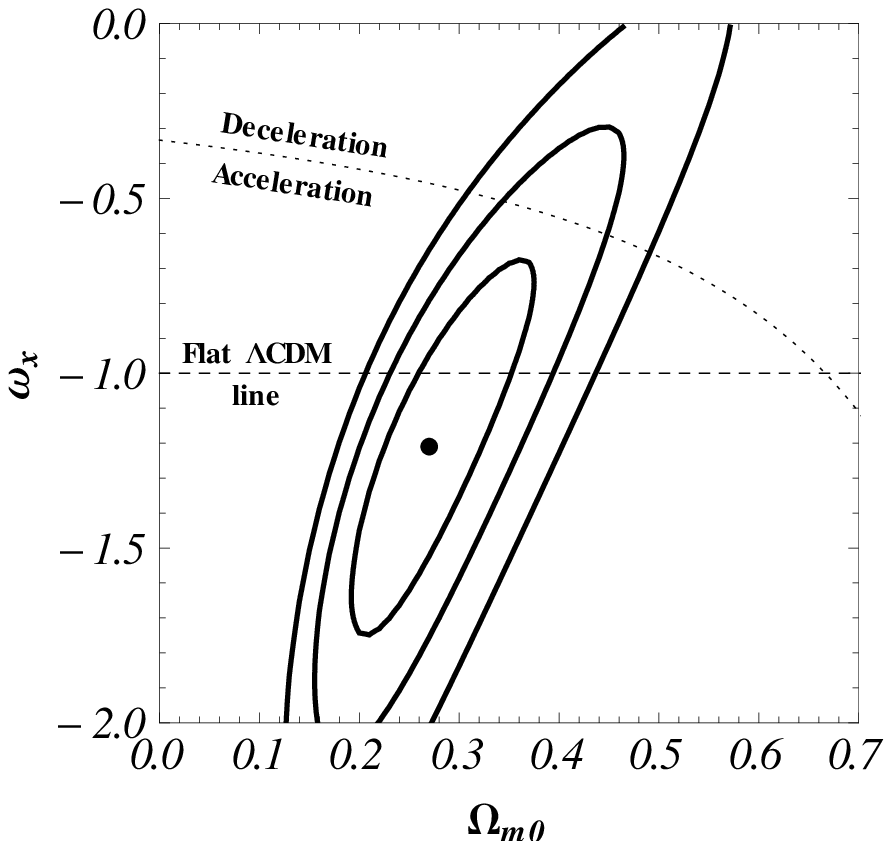}
  \includegraphics[width=79mm]{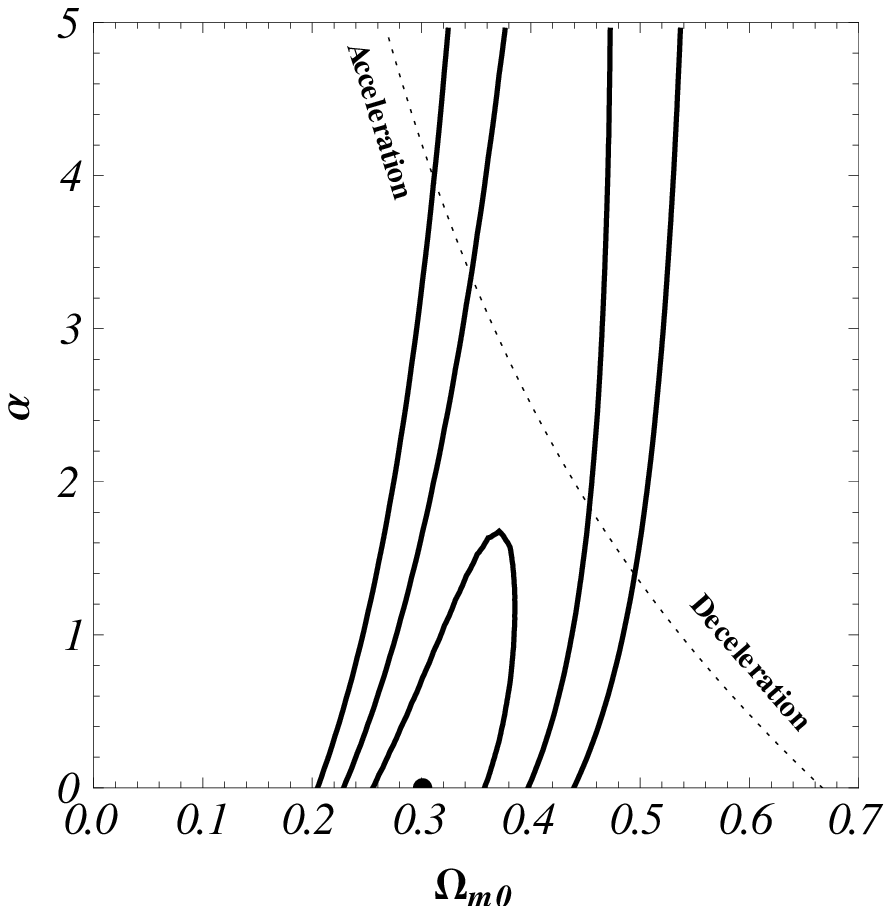}  
\caption{
1, 2, and 3 $\sigma$ constraint contours from the BAO data. Filled 
circles denote likelihood maxima. The top left plot is for the 
$\Lambda$CDM model, the top right one is for the XCDM parametrization, 
and the bottom plot is for the $\phi$CDM model. For quantitative 
details see Table \ref{tab:results-1}.
} \label{fig:allonlybao}
\end{figure}

We use measurements of the acoustic parameter $A(z)$ from \cite{blake11}, 
where the theoretically-predicted $A_{\rm th}(z)$ is given in Eq.\ (5) 
of \cite{eis05},
\begin{equation}
\label{eq:A_z}
A_{\rm th}(z)=\frac{100~D_V(z)~\sqrt{\Omega_mh^2}}{z}.
\end{equation}
Using Eqs.\ ({\ref{eq:D_V}})---({\ref{eq:A_z}}) we have 
\begin{equation}
\label{eq:A_z2}
A_{\rm th}(z)=\sqrt{\Omega_m} \left[\frac{y^2(z)}{z^2 E(z)}\right]^{1/3},
\end{equation}
which is $h$ independent and where $E(z)$ is defined in 
Sec.\ {\ref{equations}}. 

Using the WiggleZ $A_{\rm obs}(z)$ data from Table 3 of \cite{blake11}, 
we compute
\begin{equation}
\chi_{A_z}^2(\textbf{p}) = \Delta{\textbf{A}}^T ({\rm C}_{A_z})^{-1} 
    \Delta{\textbf{A}}.
\end{equation}
Here $\Delta {\textbf{A}}$ is a vector consisting of differences
 $\Delta {A_i} = A_{\rm th}(z_i;\textbf{p}) - A_{\rm obs}(z_i)$ and 
$({\rm C}_{Az})^{-1}$ is the inverse of the 3 by 3 covariance matrix 
given in Table 3 of \cite{blake11}. 

We also use the 6dFGS and SDSS data, three measurements from 
\citet{beutler2011} and \citet{Percival2010}, listed in \cite{blake11}.
In this case the distilled parameter 
\begin{equation}
\label{eq:dz}
d_{\rm th}(z)=\frac{r_s(z_d)}{D_V(z)},
\end{equation}
where $r_s(z_d)$ is the sound horizon at the drag epoch, is given in 
Eq.\ (6) of \cite{eis98}. The correlation coefficients for this case 
are also given in Table 3 of \cite{blake11}. Using the covariance 
matrix we define
\begin{equation}
\label{eq:chiBAOdz}
\chi_{d_z}^2(h,\textbf{p}) = \Delta{\textbf{d}}^T ({\rm C}_{d_z})^{-1}
   \Delta{\textbf{d}}
\end{equation} 
where $\Delta {\textbf{d}}$ is a vector consisting of differences
$\Delta {d_i} = d_{\rm th}(z_i;h,\textbf{p}) - d_{\rm obs}(z_i)$ and 
${\rm C}_{d_z}$ is the the covariance matrix \citep{blake11}.
We then marginalize over a flat prior for $H_0$ to get
\begin{equation}
\label{eq:chiBAOdz2}
\chi_{d_z}^2(\textbf{p}) = -2~\mathrm{ln}\left[
\int^\infty_0{e^{-\chi_{d_z}^2(h,\textbf{p})/2}dh}\right].
\end{equation}

Since $\chi_{A_z}^2(\textbf{p})$ and $\chi_{d_z}^2(\textbf{p})$ correspond
to independent data, the combined BAO data
\begin{equation}
\chi_{BAO}^2(\textbf{p})=\chi_{A_z}^2(\textbf{p})+ \chi_{d_z}^2(\textbf{p}) .
\end{equation}
We can maximize the likelihood by minimizing $\chi_{BAO}^2(\textbf{p})$ 
with respect to the model parameters $\textbf{p}$ to get best-fit
parameter values $\mathbf{p_0}$ and constraint contours. Figure 
\ref{fig:allonlybao} show the constraints from the BAO data on the 
three dark energy models we consider here. The XCDM parametrization 
constraints shown in this figure are in good agreement with those
shown in Fig.\ 13 of \cite{blake11}. The constraints shown in the 
other two panels of Fig.\ \ref{fig:allonlybao} have not previously 
been computed. Comparing to the $H(z)$ and SNIa constraint 
contours of Figs.\ \ref{fig:LCDM_Hz}---\ref{fig:SNEIA-lxCDM}, we see 
that the BAO contours are also very elongated, although largely orthogonal 
to the $H(z)$ and SNIa ones. Consequently, a joint analysis of these data
will result in significantly tighter constraints than those derived
using any one of these data sets.

\section{Joint constraints}
\label{Joint}

\begin{figure}[t]
\centering
  \includegraphics[angle=0,width=80mm]{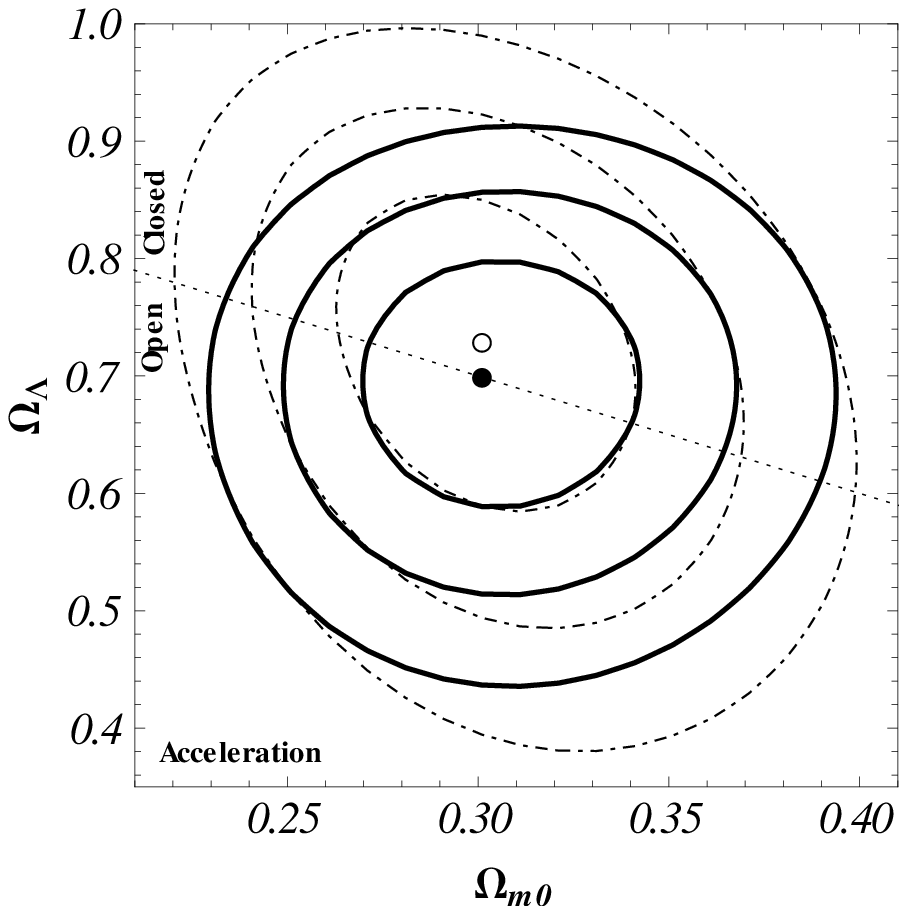}
  \includegraphics[angle=0,width=80mm]{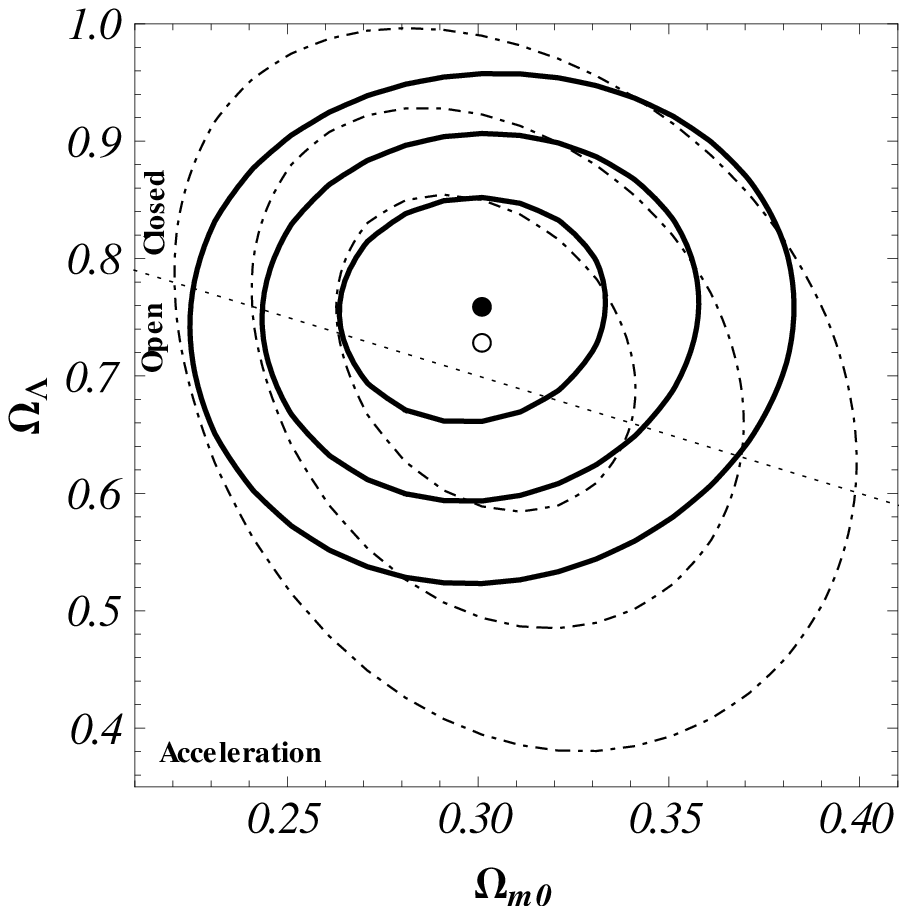}
\caption{
Thick solid (dot-dashed) lines are 1, 2, and 3 $\sigma$ constraint
contours for the $\Lambda$CDM model from a joint analysis of the BAO
and SNIa (with systematic errors) data, with (without) the $H(z)$ data. 
The full (empty) circle marks the best-fit point determined from the 
joint analysis with (without) the $H(z)$ data. The dotted sloping line 
corresponds to spatially-flat $\Lambda$CDM models. In the left panel 
we use the $H_0$ = 68 $\pm$ 2.8 km s$^{-1}$ Mpc$^{-1}$ prior while 
the right panel is for the $H_0$ = 73.8 $\pm$ 2.4 km s$^{-1}$ Mpc$^{-1}$ 
case. For quantitative details see Table \ref{tab:results-2}.
}
\label{fig:LCDM_com}
\end{figure}

\begin{figure}[t]
\centering
  \includegraphics[angle=0,width=80mm]{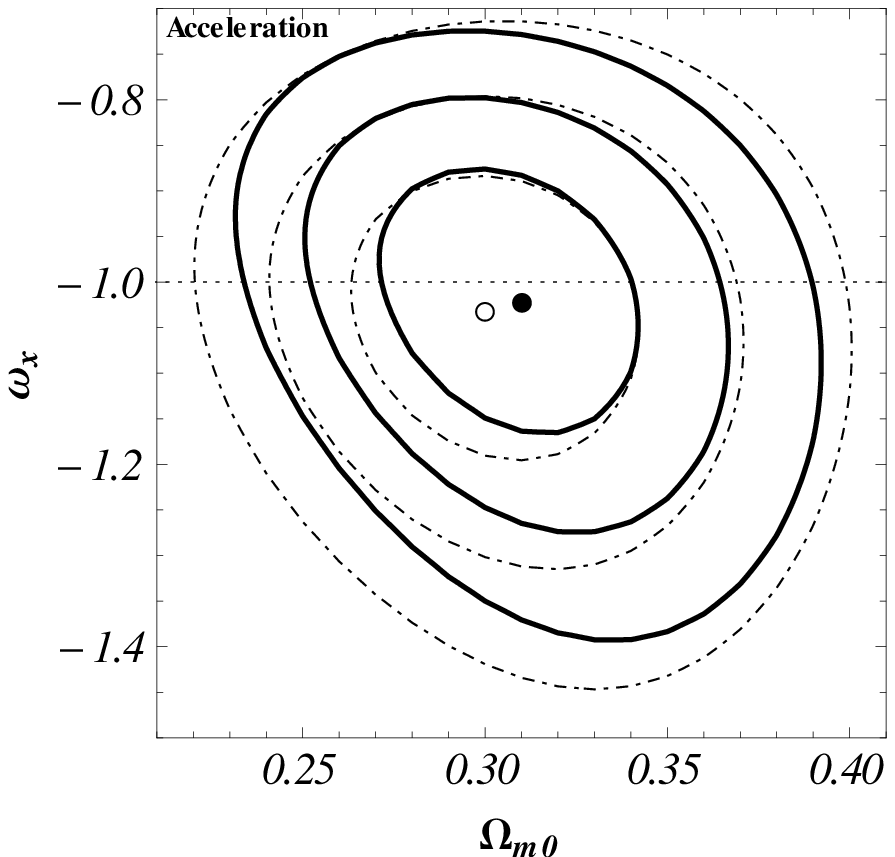}
  \includegraphics[angle=0,width=80mm]{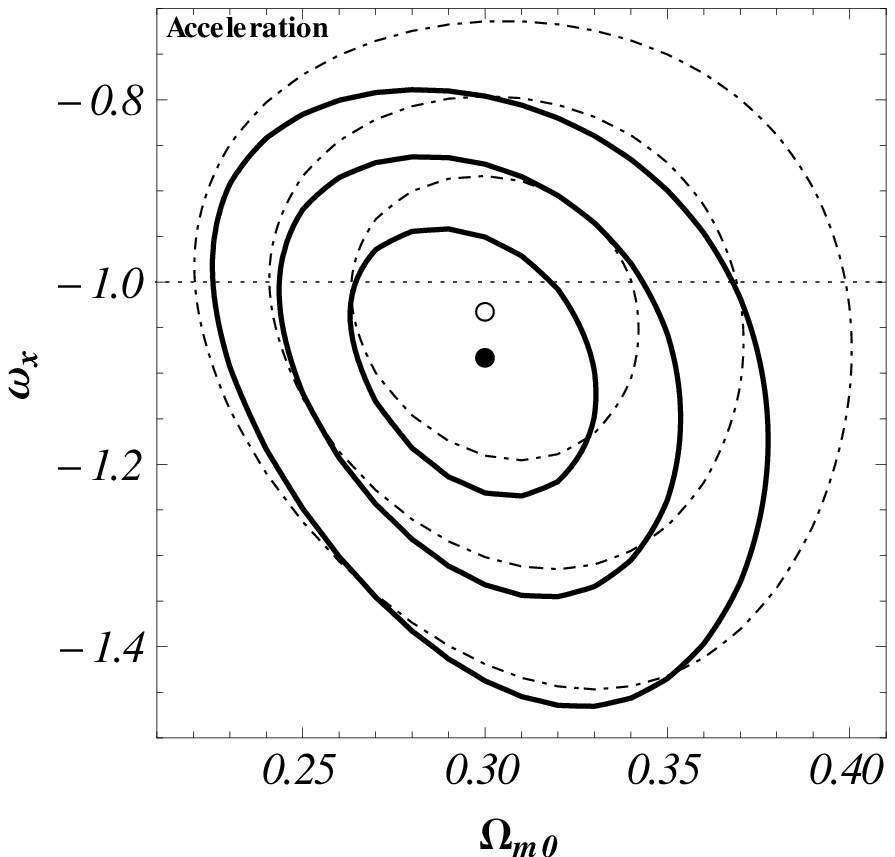}
 \caption{
Thick solid (dot-dashed) lines are 1, 2, and 3 $\sigma$ constraint
contours for the XCDM parametrization from a joint analysis of the BAO
and SNIa (with systematic errors) data, with (without) the $H(z)$ data. 
The full (empty) circle marks the best-fit point determined from the 
joint analysis with (without) the $H(z)$ data. The dotted horizontal 
line at $\omega_{\rm X} =-1$  corresponds to spatially-flat $\Lambda$CDM 
models. In the left panel we use the $H_0$ = 68 $\pm$ 2.8 km s$^{-1}$ 
Mpc$^{-1}$ prior while the right panel is for the $H_0$ = 73.8 
$\pm$ 2.4 km s$^{-1}$ Mpc$^{-1}$ case. For quantitative details see 
Table \ref{tab:results-2}.
} \label{fig:XCDM_com}
\end{figure}

To constrain cosmological parameters from a joint analysis of the $H(z)$,
SNIa, and BAO data we compute  
\begin{equation}
\chi^2(\textbf{p}) = \chi_{H}^2(\textbf{p}) + \chi_{SN}^2(\textbf{p})
      + \chi_{BAO}^2(\textbf{p})
\end{equation}
for each of the three cosmological models considered here. We minimize 
$\chi^2(\textbf{p})$ with respect to model parameters $\textbf{p}$ to 
get best-fit parameter values $\mathbf{p_0}$ and constraint contours.

\begin{figure}[t]
\centering
  \includegraphics[angle=0,width=80mm]{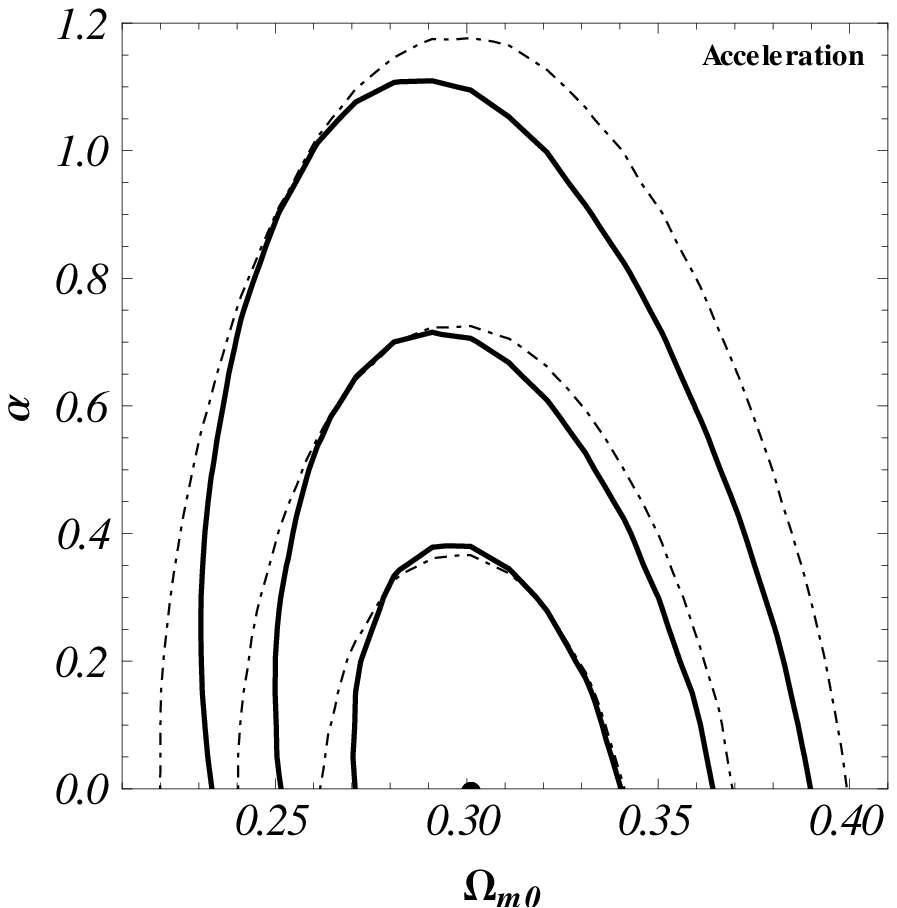}
  \includegraphics[angle=0,width=80mm]{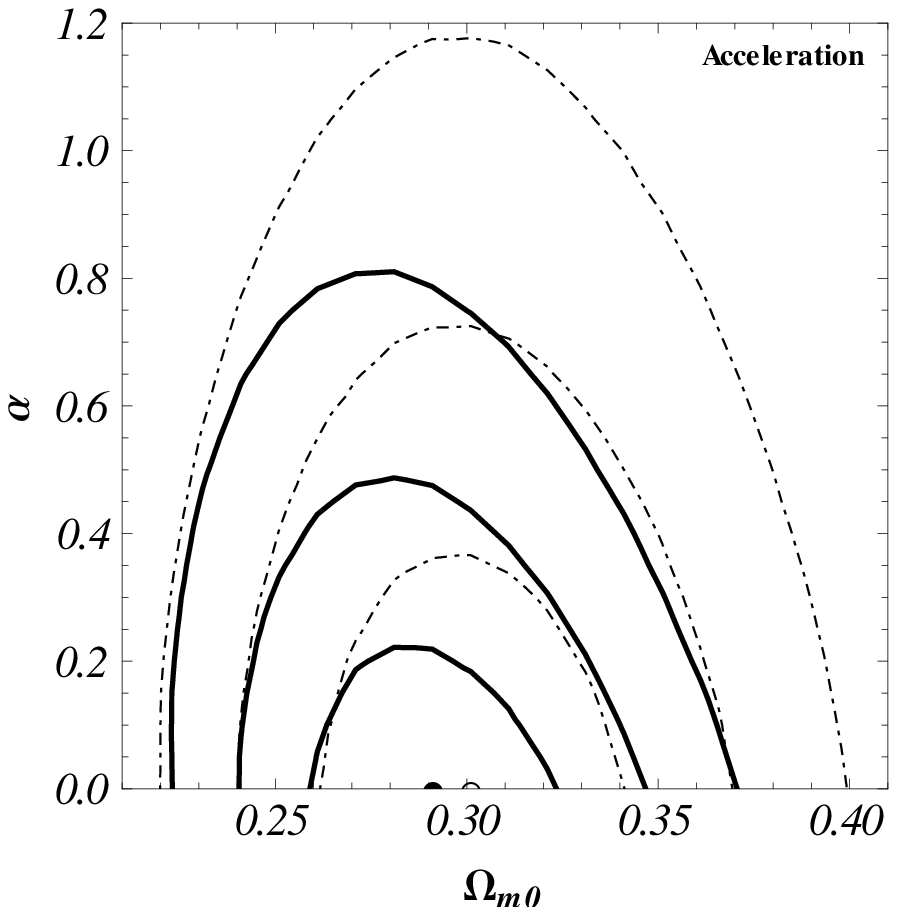}

 \caption{
Thick solid (dot-dashed) lines are 1, 2, and 3 $\sigma$ constraint
contours for the $\phi$CDM model from a joint analysis of the BAO
and SNIa (with systematic errors) data, with (without) the $H(z)$ 
data. The full (empty) circle marks the best-fit point determined
from the joint analysis with (without) the $H(z)$ data (in the left
panel the full and empty circles overlap). The $\alpha = 0$
horizontal axes correspond to spatially-flat $\Lambda$CDM models.
In the left panel we use the $H_0$ = 68 $\pm$ 2.8 km s$^{-1}$ Mpc$^{-1}$ 
prior while the right panel is for the $H_0$ = 73.8 $\pm$ 2.4 km s$^{-1}$ 
Mpc$^{-1}$ case. For quantitative details see Table \ref{tab:results-2}.
}
\label{fig:phiCDM_com}
\end{figure}

\begin{table}[htb]
\begin{center}
\begin{tabular}{ccccccc} 
\hline\hline 
{} &  
\multicolumn{2}{c}{$H(z)$+BAO}
&\multicolumn{2}{c}{$H(z)$+SNIa+BAO}
&\multicolumn{2}{c}{SNIa+BAO}\\
Model and prior & $\chi^2_{\rm min}$ & B.F.P & $\chi^2_{\rm min}$ & B.F.P & $\chi^2_{\rm min}$ & B.F.P\\ 
\hline \hline 
$\Lambda$CDM & \multirow{2}{*}{$20.7$} & $\Omega_{m0}=0.31$ & \multirow{2}{*}{566} & $\Omega_{m0}=0.30$ & \multirow{4}{*}{$551$} & \multirow{2}{*}{$\Omega_{m0}=0.30$} \\ 
$h = 0.68 \pm 0.028$ &  & $\Omega_{\Lambda}=0.68$ &  & $\Omega_{\Lambda}=0.70$ &  & {} \\ 
\cline{1-5}

$\Lambda$CDM & \multirow{2}{*}{$21.0$} & $\Omega_{m0}=0.29$ & \multirow{2}{*}{567} & $\Omega_{m0}=0.30$ & \multirow{2}{*}{} & \multirow{2}{*}{$\Omega_{\Lambda}=0.73$} \\ 
$h = 0.738 \pm 0.024$ &  & $\Omega_{\Lambda}=0.79$ &  & $\Omega_{\Lambda}=0.76$ &  & \multirow{4}{*}{}\\ 
\hline
\hline

XCDM & \multirow{2}{*}{$20.7$} & $\Omega_{m0}=0.31$ & \multirow{2}{*}{566} & $\Omega_{m0}=0.31$ & \multirow{4}{*}{$551$} & \multirow{2}{*}{$\Omega_{m0}=0.30$} \\ 
$h = 0.68 \pm 0.028$ &  & $\omega_{X}=-0.99$ &  & $\omega_{X}=-1.02$ &  & {} \\
\cline{1-5}

XCDM & \multirow{2}{*}{$20.8$} & $\Omega_{m0}=0.28$ & \multirow{2}{*}{567} & $\Omega_{m0}=0.30$ & \multirow{2}{*}{} & \multirow{2}{*}{$\omega_{X}=-1.03$} \\ 
$h = 0.738 \pm 0.024$ &  & $\omega_{X}=-1.19$ &  & $\omega_{X}=-1.08$ &  & \multirow{4}{*}{}\\ 
\hline 
\hline

$\phi$CDM & \multirow{2}{*}{$20.7$} & $\Omega_{m0}=0.31$ & \multirow{2}{*}{566} & $\Omega_{m0}=0.30$ & \multirow{4}{*}{551} & \multirow{2}{*}{$\Omega_{m0}=0.30$} \\ 
$h = 0.68 \pm 0.028$ &  & $\alpha=0.05$ &  & $\alpha=0.00$ &  & {} \\
\cline{1-5}

$\phi$CDM & \multirow{2}{*}{$22.0$} & $\Omega_{m0}=0.29$ & \multirow{2}{*}{567} & $\Omega_{m0} =0.29$ & \multirow{2}{*}{} & \multirow{2}{*}{$\alpha =0.00$} \\ 
$h = 0.738 \pm 0.024$ &  & $\alpha =0.00$ &  & $\alpha =0.00$ &  & \multirow{4}{*}{}\\ 
\hline
\hline  
\end{tabular}

\caption{The minimum value of $\chi^2$ and the corresponding best fit 
points (B.F.P) which maximize the likelihood, for different combinations 
of data. The SNIa data values are for the case including systematic errors. }
\label{tab:results-2}
\end{center}
\end{table}

\begin{figure}[htb]
\centering
  \includegraphics[angle=0,width=80mm]{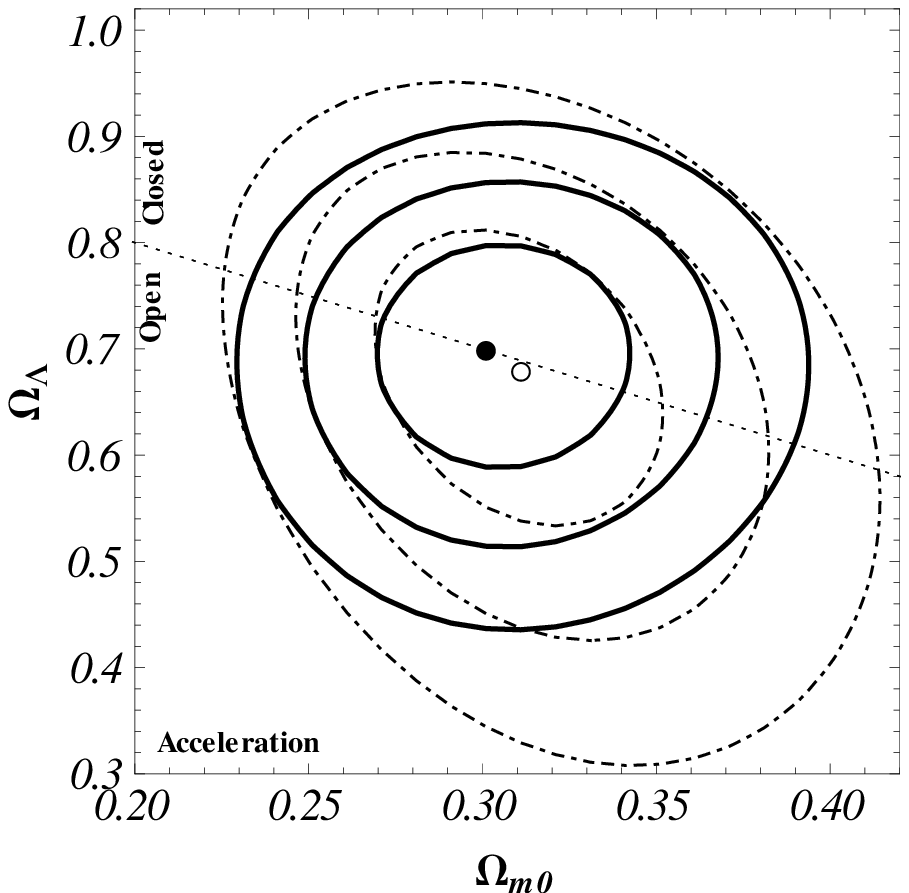}
  \includegraphics[angle=0,width=80mm]{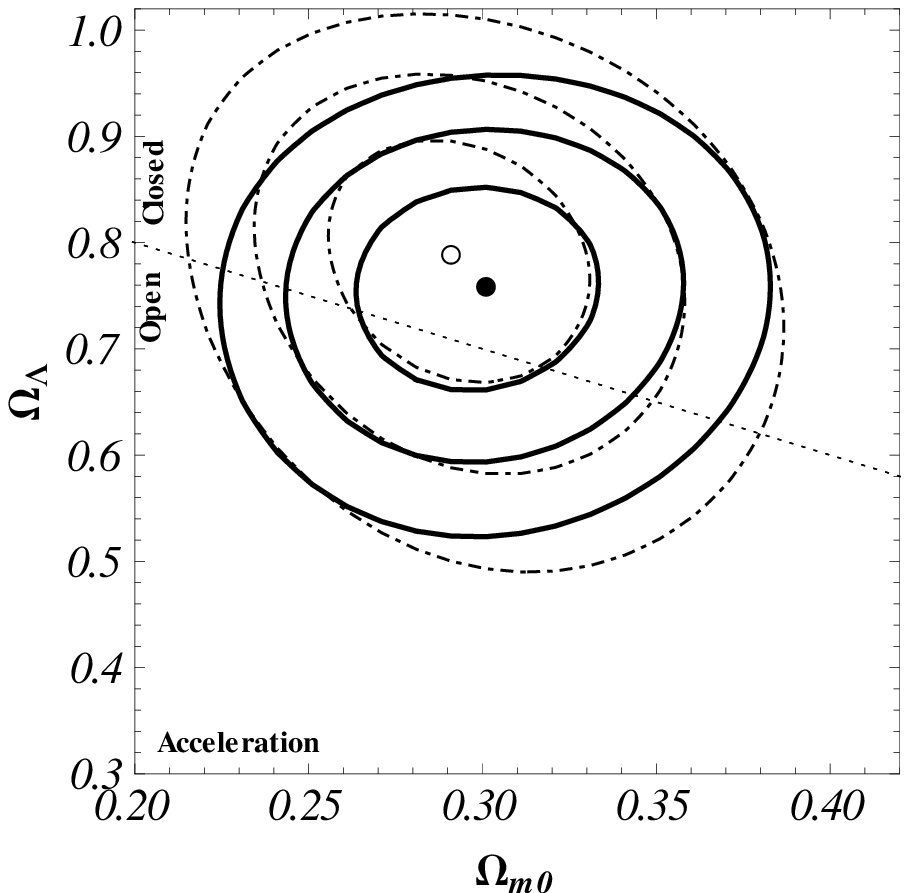}
\caption{
Thick solid (dot-dashed) lines are 1, 2, and 3 $\sigma$ constraint
contours for the $\Lambda$CDM model from a joint analysis of the BAO
and $H(z)$ data, with (without) the SNIa data. 
The full (empty) circle marks the best-fit point determined from the 
joint analysis with (without) the SNIa data. The dotted sloping line 
corresponds to spatially-flat $\Lambda$CDM models. In the left panel 
we use the $H_0$ = 68 $\pm$ 2.8 km s$^{-1}$ Mpc$^{-1}$ prior while  
the right panel is for the $H_0$ = 73.8 $\pm$ 2.4 km s$^{-1}$ Mpc$^{-1}$ 
case. For quantitative details see Table \ref{tab:results-2}.
}
\label{fig:LCDM_com2}
\end{figure}

\begin{figure}[htb]
\centering
  \includegraphics[angle=0,width=80mm]{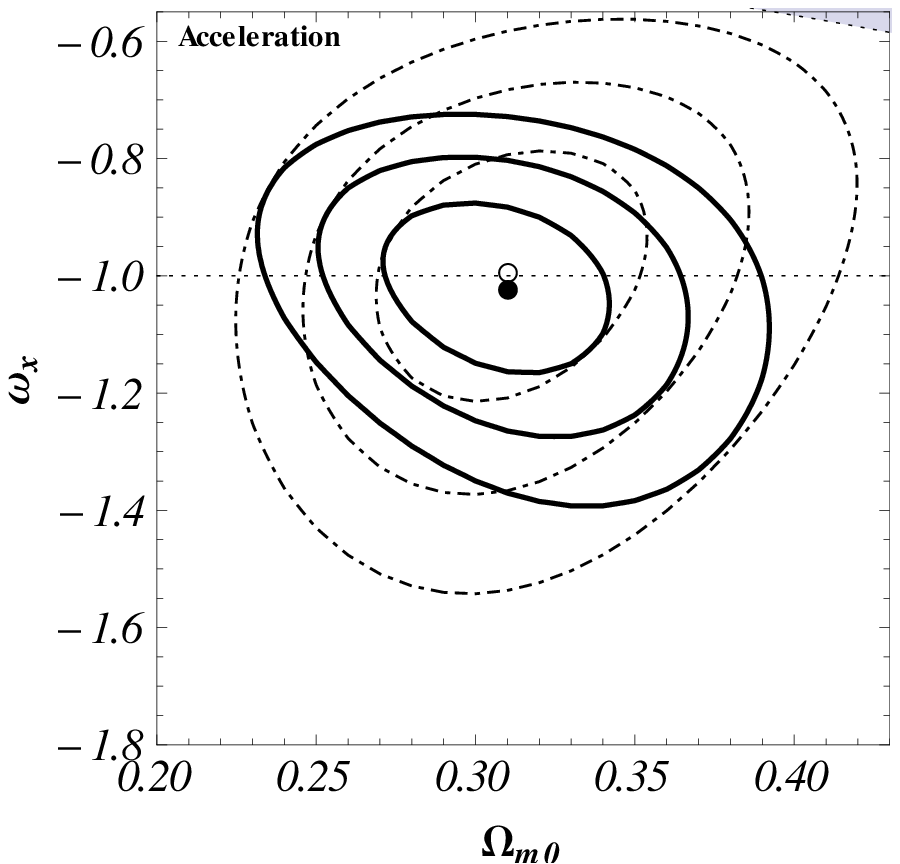}
  \includegraphics[angle=0,width=80mm]{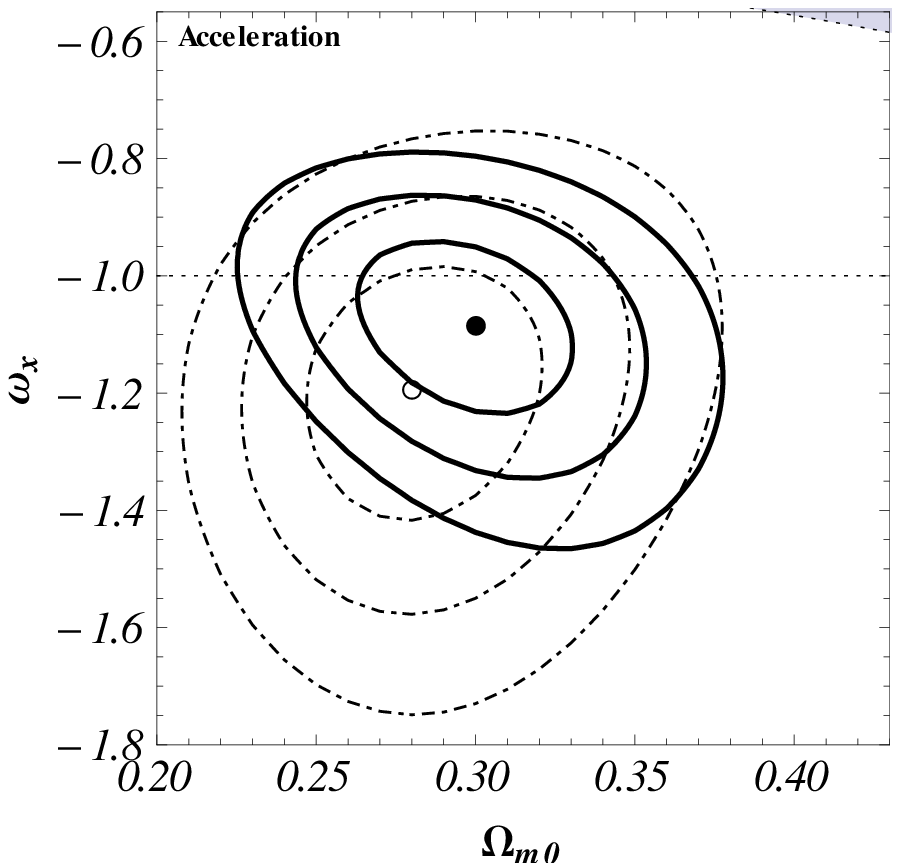}
 \caption{
Thick solid (dot-dashed) lines are 1, 2, and 3 $\sigma$ constraint
contours for the XCDM parametrization from a joint analysis of the BAO
and $H(z)$ data, with (without) the SNIa data. 
The full (empty) circle marks the best-fit point determined from the 
joint analysis with (without) the SNIa data. The dotted horizontal 
line at $\omega_{\rm X} =-1$  corresponds to spatially-flat $\Lambda$CDM 
models. In the left panel we use the $H_0$ = 68 $\pm$ 2.8 km s$^{-1}$ 
Mpc$^{-1}$ prior while the right panel is for the $H_0$ = 73.8 $\pm$ 2.4 
km s$^{-1}$ Mpc$^{-1}$ case. The shaded area in the upper right corners 
are the region of decelerating expansion. For quantitative details see
Table \ref{tab:results-2}.
} \label{fig:XCDM_com2}
\end{figure}

\begin{figure}[htb]
\centering
  \includegraphics[angle=0,width=80mm]{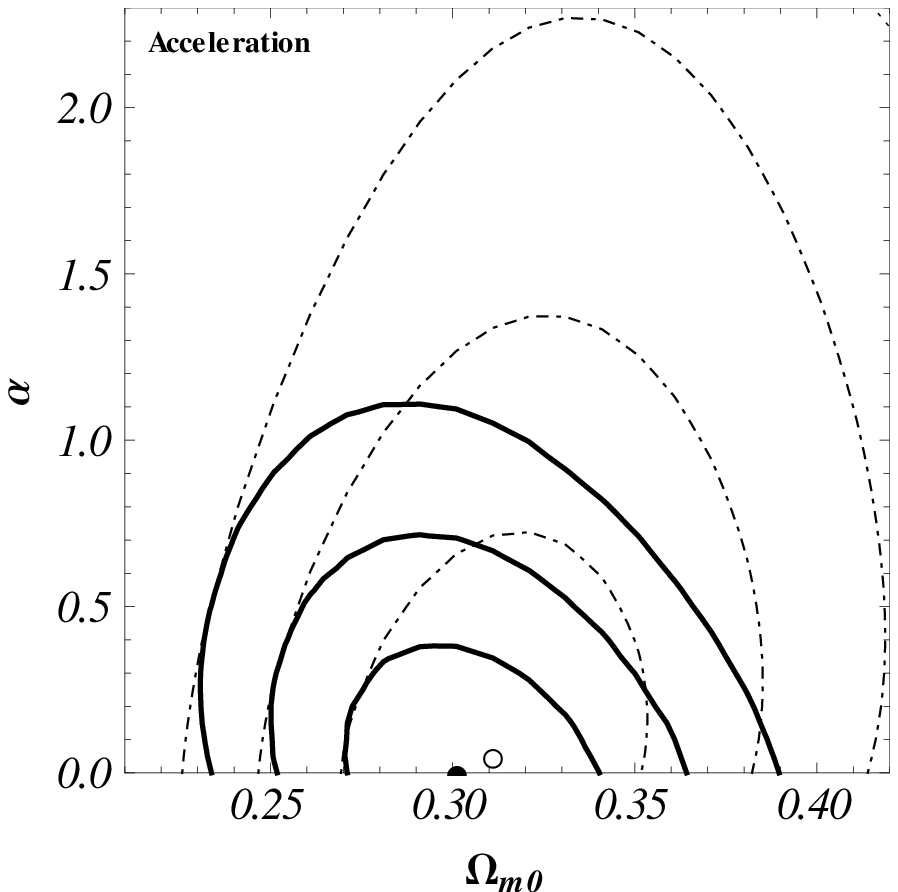}
  \includegraphics[angle=0,width=80mm]{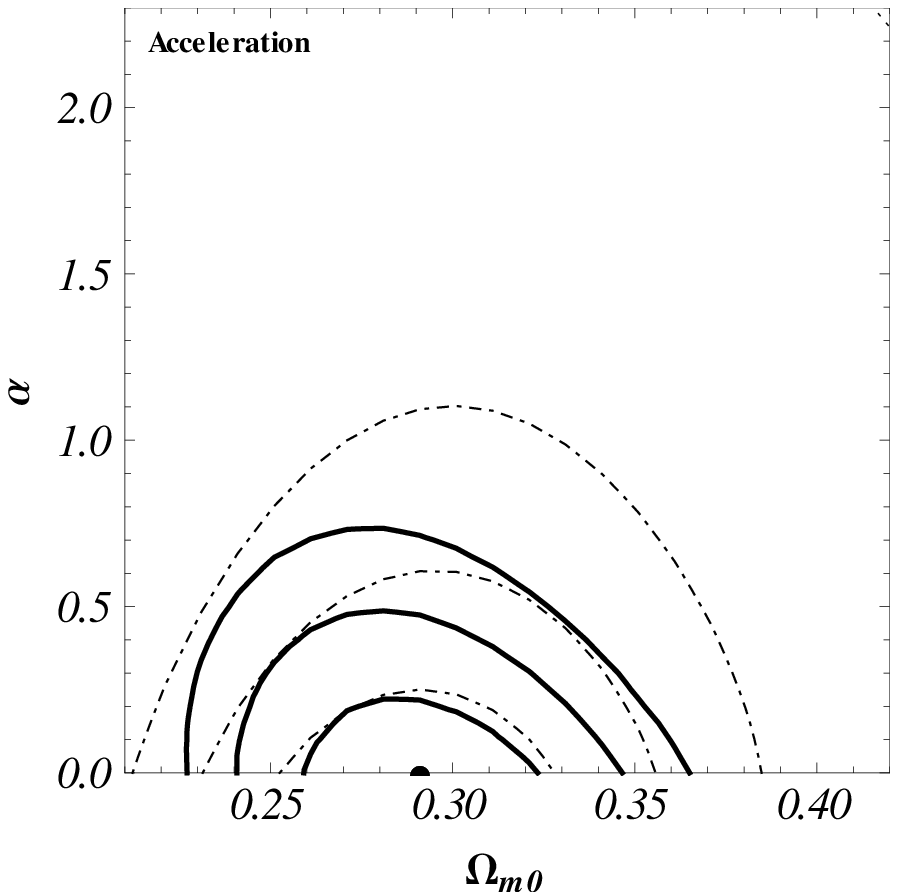}

 \caption{
Thick solid (dot-dashed) lines are 1, 2, and 3 $\sigma$ constraint
contours for the $\phi$CDM model from a joint analysis of the $H(z)$
and BAO data, with (without) the SNIa 
data. The full (empty) circle marks the best-fit point determined
from the joint analysis with (without) the SNIa data. The $\alpha = 0$
horizontal axes correspond to spatially-flat $\Lambda$CDM models.
In the left panel we use the $H_0$ = 68 $\pm$ 2.8 km s$^{-1}$ Mpc$^{-1}$ 
prior while the right panel is for the $H_0$ = 73.8 $\pm$ 2.4 km s$^{-1}$ 
Mpc$^{-1}$ case. For quantitative details see Table \ref{tab:results-2}.
}
\label{fig:phiCDM_com2}
\end{figure}

Figures \ref{fig:LCDM_com}---\ref{fig:phiCDM_com} show
constraints on the cosmological parameters for the $\Lambda$CDM and
$\phi$CDM models and the XCDM parametrization, from a joint analysis
of the BAO and SNIa data, as well as from a joint analysis of the
BAO, SNIa and $H(z)$ data. Table 3 lists information about best-fit
parameter values. Including the $H(z)$ data in the analysis 
tightens the constraints by more than one standard
deviation, in parts of the parameter spaces. 

\begin{table}[htb]
\begin{center}
\begin{tabular}{cccc}
\hline\hline
{Model and prior}	& {SNIa+BAO}	&{$H(z)$+BAO} & {$H(z)$+SNIa+BAO} \\
\hline
 {$\Lambda $CDM } & {0.25 $<$ }$\Omega_{m0}${ $<$ 0.36} &  {0.25 $<$ }$\Omega_{m0}${ $<$ 0.36} & {0.26 $<$ }$\Omega_{m0}${ $<$ 0.36} \\
 {$h=0.68 \pm 0.028$} & {0.53 $<$ }$\Omega _{\Lambda }${ $<$ 0.89} & {0.45 $<$ }$\Omega _{\Lambda }${ $<$ 0.85}  &{0.55 $<$ }$\Omega _{\Lambda }${ $<$ 0.83} \\
 \hline

 {$\Lambda $CDM} & {0.25 $<$ }$\Omega_{m0}${ $<$ 0.36} & {0.23 $<$ }$\Omega_{m0}${ $<$ 0.38} & {0.25 $<$ }$\Omega_{m0}${ $<$ 0.35} \\
 {$h = 0.738 \pm 0.024$} & {0.53 $<$ }$\Omega _{\Lambda }${ $<$ 0.89} & {0.60 $<$ }$\Omega _{\Lambda }${ $<$ 0.92} &{0.62 $<$ }$\Omega _{\Lambda }${ $<$ 0.88} \\
 \hline

 {XCDM} & {0.30 $<$ }$\Omega_{m0}${ $<$ 0.38} & {0.26 $<$ }$\Omega_{m0}${ $<$ 0.37} &  {0.29 $<$ }$\Omega_{m0}${ $<$ 0.37} \\
 {$h = 0.68 \pm 0.028$} & $-1.18 <  \omega_{\rm X} < -0.78$ & $-1.32 <  \omega_{\rm X} < -0.73$ & $-1.14 < \omega_{\rm X} < -0.78$ \\
 \hline 

 {XCDM} & {0.30 $<$ }$\Omega_{m0}${ $<$ 0.38} & {0.24 $<$ }$\Omega_{m0}${ $<$ 0.35} & {0.27 $<$ }$\Omega_{m0}${ $<$ 0.35} \\
 { $h = 0.738 \pm 0.024$} & $-1.18 < \omega_{\rm X} < -0.78$ & $-1.42 < \omega_{\rm X} < -0.88$ & $ -1.22 < \omega_{\rm X} < -0.86$ \\
  \hline

 {$\phi $CDM} & {0.25 $<$ }$\Omega_{m0}${ $<$ 0.35} & {0.25 $<$ }$\Omega_{m0}${ $<$ 0.36} & {0.26 $<$ }$\Omega_{m0}${ $<$ 0.35} \\
 {$h = 0.68 \pm 0.028$} & {0 $<$ $\alpha $ $<$ 0.54} & {0 $<$ $\alpha $ $<$ 1.01} & {0 $<$ $\alpha $ $<$ 0.54} \\
  \hline

 {$\phi $CDM} & {0.25 $<$ }$\Omega_{m0}${ $<$ 0.35} & {0.23 $<$ }$\Omega_{m0}${ $<$ 0.35} & {0.25 $<$ }$\Omega_{m0}${ $<$ 0.33} \\
 {$h = 0.738 \pm 0.024$} & {0 $<$ $\alpha $ $<$ 0.54} & {0 $<$ $\alpha $ $<$ 0.57} & {0 $<$ $\alpha $ $<$ 0.35} \\
\hline\hline
\end{tabular}

\caption{Two standard deviation bounds on cosmological
parameters using SNIa+BAO, $H(z)$+BAO and SNIa+BAO+$H(z)$ data, for 3 
different models with two different $H_0$ priors.}
\label{tab:intervals}
\end{center}
\end{table}

Adding the $H(z)$ data for the $\bar{H_{0}}\pm\sigma_{H_{0}} 
= 68 \pm 2.8$ km s$^{-1}$ Mpc$^{-1}$ prior case improved the
constraints most significantly in the $\Lambda$CDM case (by more
than 1 $\sigma$ on $\Omega_{\Lambda}$ in parts of parameter space), 
Fig.\ \ref{fig:LCDM_com}, and least significantly for the $\phi$CDM model,
Fig.\ \ref{fig:phiCDM_com}. For the case of the 
$\bar{H_{0}}\pm\sigma_{H_{0}} = 73.8 \pm 2.4$ km s$^{-1}$ Mpc$^{-1}$ prior, 
adding $H(z)$ again tightens up the constraints the most for the 
$\Lambda$CDM model (by more than 1 $\sigma$ on $\Omega_{\Lambda}$), 
Fig.\ \ref{fig:LCDM_com}, and least so for the 
XCDM parametrization, Fig.\ \ref{fig:XCDM_com}.

Figures \ref{fig:LCDM_com2}---\ref{fig:phiCDM_com2} show the constraints 
on the cosmological parameters of the three models, from a joint analysis 
of the BAO and $H(z)$ data, as well as from a joint analysis of the 
three data sets. Table 3 lists the best-fit parameter values. Comparing these 
figures to Figs.\ \ref{fig:LCDM_com}---\ref{fig:phiCDM_com} allows for a
comparison between the discriminating power of the SNIa and $H(z)$ data.  

Figure \ref{fig:LCDM_com2} shows that adding SNIa data to the $H(z)$ and BAO 
data combination for the $\bar{H_{0}}\pm\sigma_{H_{0}} = 68 \pm 2.8$ 
km s$^{-1}$ Mpc$^{-1}$ prior case tightens up the constraints by more than 
1 $\sigma$ on $\Omega_{\Lambda}$ from below, while addition of SNIa data 
for the $\bar{H_{0}}\pm\sigma_{H_{0}} = 73.8 \pm 2.4$ km s$^{-1}$ Mpc$^{-1}$ 
prior case tightens up the constraints by more than 1 $\sigma$ on 
$\Omega_{\Lambda}$ from above. Addition of SNIa data to the $H(z)$ and BAO 
combination doesn't much improve the constraints on $\Omega_{m0}$ for 
either prior.

Figures \ref{fig:LCDM_com2}---\ref{fig:phiCDM_com2} show that adding 
SNIa data to the $H(z)$ and BAO combination results in the most prominent 
effect for the XCDM case, Fig.\ \ref{fig:XCDM_com2}. Here for the 
$\bar{H_{0}}\pm\sigma_{H_{0}} = 68 \pm 2.8$ km s$^{-1}$ Mpc$^{-1}$ 
prior it tightens up the constraints by more than 1 $\sigma$ on 
$\omega_{X}$ from above and below while for the 
$\bar{H_{0}}\pm\sigma_{H_{0}} = 73.8 \pm 2.4$ km s$^{-1}$ Mpc$^{-1}$ 
prior it tightens up the constraints by more than 2 $\sigma$ on $\omega_{X}$
from below. Addition of SNIa data to the $H(z)$ and BAO combination 
doesn't  much improve the constraints on $\Omega_{m0}$ for either prior
in this case.

In the $\phi$CDM case, Fig.\ \ref{fig:phiCDM_com2}, adding SNIa data to 
$H(z)$ and BAO combination affects the constraint on $\alpha$ the most
for the $\bar{H_{0}}\pm\sigma_{H_{0}} = 68 \pm 2.8$ km s$^{-1}$ Mpc$^{-1}$ 
prior case. The effect on $\Omega_{m0}$ is little stronger than what 
happens in the $\Lambda$CDM and XCDM cases but still less than 1 $\sigma$.

Table 4 lists the two standard deviation bounds on the individual 
cosmological parameters, determined from their one-dimensional
posterior probability distributions  functions (which are derived
by marginalizing the two-dimensional likelihood over the other
cosmological parameter) for different combinations of data set. 

The constraints on the cosmological parameters that we derive from 
only the BAO and SNIa data are restrictive, but less so than those 
shown in Fig.\ 4 of \citet{Chen2011b}. This is because the new 
\citet{suzuki12} SNIa compilation data we use here is based on a more
careful accounting of the systematic errors, which have increased. 
Consequently, including the $H(z)$ data, in addition to the BAO and 
SNIa data, in the analysis, more significantly tightens 
the constraints: compare Figs.\ \ref{fig:LCDM_com}---\ref{fig:phiCDM_com}
here to Figs.\ 4---6 of \citet{Chen2011b}. We emphasize, however, that
this effect is prominent only in some parts of the parameter spaces. 

\section{Conclusion}
\label{summary}

In summary, the results of a joint analysis of the $H(z)$, BAO, and SNIa 
data are very consistent with the predictions of a spatially-flat 
cosmological model with energy budget dominated by a time-independent 
cosmological constant, the standard $\Lambda$CDM model. However, the 
data are not yet good enough to strongly rule out slowly-evolving dark 
energy density. More, and better quality, data are needed to discriminate
between constant and slowly-evolving dark energy density.

It is probably quite significant that current $H(z)$ data constraints
are almost as restrictive as those from SNIa data. The acquisition 
of $H(z)$ data has been an interesting backwater of cosmology for the 
last few years. We hope that our results will help promote more 
interest in this exciting area. Since the $H(z)$ technique has not 
been as much studied as, say, the SNIa apparent magnitude technique, 
a little more effort in the $H(z)$ area is likely to lead to very
useful results.

\acknowledgments

We thank Chris Blake, Michele Moresco, Larry Weaver, and 
Shawn Westmoreland for useful discussions and helpful advice. 
We are grateful to the referee for a very detailed and prompt
report that helped us improve our manuscript.
This work was supported in part by DOE grant DEFG03-99EP41093 
and NSF grant AST-1109275.


\end{document}